\documentclass[a4paper,11pt]{article}
\pdfoutput=1 

\usepackage[utf8]{inputenc}
\usepackage[T1]{fontenc}
\usepackage[margin=1in]{geometry}
\usepackage{parskip}
\usepackage{booktabs,multirow}
\usepackage{subcaption}
\usepackage[margin=1cm]{caption}
\usepackage{mathtools,amssymb,braket,dsfont}
\usepackage{graphicx}
\usepackage{xcolor}
\usepackage{cite}

\usepackage[final]{hyperref}
\hypersetup{
	colorlinks=true,
	linkcolor=blue,
	citecolor=blue,
	filecolor=magenta,
	urlcolor=blue
}

\newcommand{\e}{\mathrm{e}}
\newcommand{\fli}{\alpha}
\newcommand{\cpasym}[2]{\varepsilon_{N_{#2}}^{#1}}
\newcommand{\cpasyms}[2]{\varepsilon_{\widetilde{N}_{#2}}^{#1}}
\newcommand{\decK}[2]{K_{N_{#2}}^{#1}}
\newcommand{\diag}{\mathrm{diag}}
\newcommand{\pmatr}[1]{\begin{pmatrix} #1 \end{pmatrix}}
\newcommand{\overbar}[1]{\mkern 1.5mu\overline{\mkern-1.5mu#1\mkern-1.5mu}\mkern 1.5mu}
\newcommand{\BesselK}[1]{\mathcal{K}_{#1}(z)}

\begin{document}

\hfill {\small IFIC/19-21,\quad FTUV-19-0416}\\[0.5cm]

\begin{center}
	{\Large \bf \boldmath Leptogenesis in $\Delta(27)$ with a Universal Texture Zero} \\[1cm]
	{\large
	Fredrik Bj\"{o}rkeroth$^\star$\footnote{fredrik.bjorkeroth@lnf.infn.it}, 
 	Ivo de Medeiros Varzielas$^\dagger$\footnote{ivo.de@udo.edu}, 
	M.L. L\'{o}pez-Ib\'{a}\~{n}ez$^\ddagger$\footnote{maloi2@uv.es}, 
 	Aurora Melis$^\ddagger$\footnote{aurora.melis@uv.es},
 	\'{O}scar Vives$^\ddagger$\footnote{oscar.vives@uv.es}
 	}\\[20pt]
	{\small
	$^\star$ INFN, Laboratori Nazionali di Frascati, C.P. 13, 100044 Frascati, Italy \\[5pt]
	$^\dagger$  CFTP, Departamento de Física, Instituto Superior T\'{e}cnico, Universidade de Lisboa, \\ Avenida Rovisco Pais 1, 1049 Lisboa, Portugal \\[5pt]
	$^\ddagger$ Departament de F\'{i}sica T\`{e}orica, Universitat de Val\`{e}ncia \& IFIC, Universitat de Val\`{e}ncia \& CSIC, \\ Dr. Moliner 50, E-46100 Burjassot (Val\`{e}ncia), Spain
	}
\end{center}
\vspace*{1cm}

\begin{abstract}
\noindent 
We investigate the possibility of viable leptogenesis in an appealing $\Delta(27)$ model with a universal texture zero in the (1,1) entry. 
The model accommodates the mass spectrum, mixing and CP phases for both quarks and leptons and allows for grand unification. 
Flavoured Boltzmann equations for the lepton asymmetries are solved numerically, taking into account both $N_1$ and $N_2$ right-handed neutrino decays.
The $N_1$-dominated scenario is successful and the most natural option for the model, with $M_1 \in [10^9, 10^{12}]$ GeV, and $M_1/M_2 \in [0.002, 0.1]$, which constrains the parameter space of the underlying model and yields lower bounds on the respective Yukawa couplings.
Viable leptogenesis is also possible in the $N_2$-dominated scenario, with the asymmetry in the electron flavour protected from $N_1$ washout by the texture zero.
However, this occurs in a region of parameter space which has a stronger mass hierarchy $M_1/M_2 < 0.002 $, and $M_2$ relatively close to $M_3$, which is not a natural expectation of the $\Delta(27)$ model.
\end{abstract}

\newpage

\section{Introduction}
\label{sec:Intro}

The Standard Model (SM) has been experimentally confirmed as the correct description of Nature, with excellent precision, up to scales of $\mathcal{O}$(TeV). 
Nevertheless we know that the SM is not a complete theory.
It includes a host of free parameters, the majority of which relate to the Yukawa sector, into whose origin and nature the SM offers no insight.
This is despite obvious indications of internal structure, such as large mass hierarchies between generations of fermions, and small CKM mixing.
It is also unclear as to how the SM should be extended to account for massive neutrinos and lepton mixing.
The combined questions of charged fermion hierarchies and the CKM and PMNS mixing patterns is typically referred to as the flavour puzzle.
Moreover, the SM fails to accommodate several observational facts in cosmology. 
It lacks dark matter and inflaton candidates, has no explanation for dark energy, and does not account for the baryon asymmetry of the Universe (BAU).

Among these cosmological issues, perhaps the BAU is the most distressing one. 
The SM is (nearly) symmetric in particles and anti-particles; despite this, no evidence of the presence of primordial anti-matter in our observable universe has been found so far.
The BAU, that is, the difference between the baryon $n_B$ and antibaryon $\overbar{n}_B$ number densities, is measured with respect to the entropy density $s$ to be
\begin{equation}
    Y_B =\frac{n_B-\overbar{n}_B}{s} = (0.87 \pm 0.01)\times 10^{-10}.
\label{eq:BAU}
\end{equation}
Although the SM includes all the necessary ingredients to generate this BAU dynamically \cite{Sakharov:1967dj}, namely, CP violation in the CKM matrix, $B$ violation through sphaleron interactions, and out-of-equilibrium processes in the electroweak phase transition, the asymmetry obtained in the SM is too small by orders of magnitude \cite{Kuzmin:1985mm}.

It is well-known that extending the SM by several heavy right-handed (RH) neutrinos can yield a BAU via leptogenesis \cite{Fukugita:1986hr}.
Lepton number-violating decays of the RH neutrinos, some portion of which occur out of equilibrium, produce a lepton asymmetry.
This is partially converted into a baryon asymmetry by sphaleron interactions, which are efficient above the electroweak scale.
Heavy RH neutrinos simultaneously provide a natural answer to the smallness of left-handed (LH) neutrino masses via the seesaw mechanism.

It is interesting to note that since RH neutrinos are SM singlets, leptogenesis links the resolution of the BAU with their Yukawa couplings, and thus connects with the flavour puzzle.
If seesaw is indeed the origin of light neutrino masses, then qualitatively leptogenesis is unavoidable. 
Whether it accurately reproduces the observed BAU becomes a quantitative question for a given spectrum of RH neutrinos and their interactions with SM particles.
Remarkably, the original (and arguably simplest) model of leptogenesis requires a RH neutrino scale $M \gtrsim 10^9 $ GeV, which closely corresponds to the ``natural'' seesaw scale.

The flavour sector of the SM, including lepton mixing, comprises 22 (20) physical parameters, assuming neutrinos are Majorana (Dirac) particles.
A popular approach to relate these parameters, and reduce the effective number of degrees of freedom in the SM, is that of spontaneously broken flavour (or family) symmetries.
Non-Abelian discrete symmetries have been especially successful, able to simultaneously describe charged lepton and neutrino parameters, and in several cases, also the quark sector \cite{Blum:2007jz, Hagedorn:2012pg, Holthausen:2013vba, Ishimori:2014nxa, Yao:2015dwa, Varzielas:2016zuo, Lu:2018oxc, Hagedorn:2018gpw, Lu:2019gqp}.
A very appealing $\Delta(27)$ model was introduced in \cite{deMedeirosVarzielas:2017sdv}, consistent with an underlying $SO(10)$ grand unified theory (GUT). 
The family symmetry leads to a predictive structure with a universal texture zero (UTZ) for all fermion mass structures, including the effective neutrinos after seesaw. 
The family symmetry is also responsible for controlling flavour-violating processes, which are sufficiently suppressed for certain regions of the model parameter space, as shown in \cite{deMedeirosVarzielas:2018vab}.
A complete model ought to account for the observed BAU, which provides an additional constraints on its parameters.
In particular, as we shall see in this analysis, matching to the observed BAU allows us to constrain the otherwise unknown parameters of the RH neutrino sector.  

The paper is organized as follows. 
Section~\ref{sec:Model} summarizes the main features of the model, originally presented in \cite{deMedeirosVarzielas:2017sdv}. 
The seesaw implementation is described in Section~\ref{sec:SeeSaw}, explaining the existing UTZ result in an elegant new way based on rank-one matrices (described in more detail in Appendix \ref{app:seesawrank1}).
In Section~\ref{sec:Leptogenesis} we write down the Boltzmann equations describing the evolution of the neutrino and asymmetry densities; this is supplemented by Appendix \ref{app:boltzmann}.
Section~\ref{sec:Analysis} presents the results and our analysis. 
We conclude in Section~\ref{sec:Conclusions}.
Appendices \ref{app:seesawrank1} and \ref{app:LpGnparam} provide additional insight into the model and leptogenesis within it.

\section{Overview of the \texorpdfstring{$\boldmath{\Delta(27)}$}{Delta(27)} Model}
\label{sec:Model}

In this section we review the model introduced in \cite{deMedeirosVarzielas:2017sdv}.
Given that we are interested in leptogenesis, 
we focus on the lepton sector, where SM fermions are contained in superfields $L$ (lepton $SU(2)$ doublets) and $e^c$, $N^c$ (following conventional notation, conjugates of the RH charged leptons and neutrinos, respectively).
The field content and their transformation properties under the $\mathcal{G}_{f} = \Delta(27)\times \mathbb{Z}_N$ flavour group are given in Table~\ref{tab:D27pc}.

\begin{table}[ht]
\centering 
\begin{tabular}{c ccccccccccc}
\toprule
Field & $L$ & $ e^c $ & $ N^c$ & $H_{u,d}$ & $\Sigma$ &
$S$ & $\phi_c$ & $\phi_b$ & $\phi_a$ & $\phi$ & $\phi_X$ \\ 
\midrule
$\Delta(27)$ & $\bf 3$ & $\bf 3$ & $\bf 3$ & $\bf 1$ & $\bf 1$  & $\bf 1$ & $\overbar{\bf 3}$ & $\overbar{\bf 3}$ & $\overbar{\bf 3}$ & $\overbar{\bf 3}$  & \bf 3\\
$\mathbb{Z}_N$    & 0 & 0 & 0 & 0 & 2 & -1 & 0 & -1 & 2 & 0 & $x$ \\
\bottomrule
\end{tabular}
\caption{\small Representations of superfields under the flavour symmetry $\mathcal{G}_{f} = \Delta(27)\times \mathbb{Z}_N$.}
\label{tab:D27pc}
\end{table}

The superpotential that generates the Yukawa structures at leading order for Dirac fermions is
\begin{equation}
\begin{aligned}
    \mathcal{W}_Y & = L_{i} e^c_j H_d 
        \left[\frac{g^\e_c}{\Lambda^2} \phi_c^i \phi_c^j 
        + \frac{g^\e_b}{\Lambda^3} \phi_b^i \phi_b^j \Sigma 
        + \frac{g^\e_a}{\Lambda^3}(\phi_a^i \phi_b^j +\phi_b^i \phi_a^j ) S \right] \\
        &\quad 
        + L_{i} N^c_j H_u 
        \left[\frac{g^\nu_c}{\Lambda^2} \phi_c^i \phi_c^j 
        + \frac{g^\nu_b}{\Lambda^3} \phi_b^i \phi_b^j \Sigma 
        + \frac{g^\nu_a}{\Lambda^3} (\phi_a^i \phi_b^j + \phi_b^i \phi_a^j ) S \right] .
\end{aligned}
\label{eq:Wf}
\end{equation}
where $i,j = (1,2,3)$ lower (upper) indices denote the $\Delta(27)$ triplets (anti-triplets).
Non-renormalizable terms are suppressed by messenger masses, which are in general different \cite{deMedeirosVarzielas:2017sdv}; they are denoted here by a common scale $\Lambda$, with variations in messenger masses contained in an arbitrary coupling $g$ for each term.
The superpotential responsible for RH neutrino Majorana masses is
\begin{equation}
    \mathcal{W}_N = N_i^c N_j^c 
    \left[\frac{g^N_c}{\Lambda} \phi^i \phi^j 
        + \frac{g^N_b}{\Lambda^4} \phi_b^i \phi_b^j (\phi^k \phi^k \phi^k_a) 
        + \frac{g^N_a}{\Lambda^4} (\phi_a^i \phi_b^j + \phi_b^i \phi_a^j) (\phi^k\phi^k\phi^k_b) \right] .
\label{eq:WN}
\end{equation}

The $\phi$ fields are flavons that break $\Delta(27)$ and provide the structure of the mass matrices, with the vacuum alignment
\begin{equation}
    \braket{\phi_c} = v_c \pmatr{0\\ 0\\ 1} \propto \braket{\phi} ,\qquad
    \braket{\phi_b} = \frac{v_b}{\sqrt{2}}  \pmatr{0 \\ 1 \\ 1} ,\qquad
    \braket{\phi_a} = \frac{v_a}{\sqrt{3}}  \pmatr{1 \\ 1 \\  -1}.
\end{equation}

The core prediction of the model is universal complex-symmetric mass matrices with the UTZ in the (1,1) entry, of the form
\begin{equation}
    M = \pmatr{
        0 & a & a \\
        a & b+2 a & b\\
        a & b & c+b -2 a
    }, 
\label{eq:Mgeneric}
\end{equation}
for some complex $a$, $b$, $c$.
Assuming a strong hierarchy $ a < b < c$, the eigenvalues are approximately given by $|a^2/b|$, $|b|$ and $|c|$.
This applies in particular to the Dirac and Majorana mass matrices. 
Up to $\mathcal{O}(1)$ coefficients, they yield the following hierarchies between families:
\begin{equation}
    Y_{\e,\nu}  \sim  y_c^{\e,\nu} 
    \begin{pmatrix}
        0 & \epsilon_{\e,\nu}^3 & \epsilon_{\e,\nu}^3 \\
        \epsilon_{\e,\nu}^3 & \epsilon_{\e,\nu}^2 & \epsilon_{\e,\nu}^2\\
        \epsilon_{\e,\nu}^3 & \epsilon_{\e,\nu}^2 & 1\\ 
    \end{pmatrix} ,
    \qquad
    M_N  \sim  M_c \begin{pmatrix}
        0 & \epsilon_N^2 & \epsilon_N^2 \\
        \epsilon_N^2 & \epsilon_N^2 & \epsilon_N^2\\
        \epsilon_N^2 & \epsilon_N^2 & 1\\ 
    \end{pmatrix} ,
\label{eq:Meps}
\end{equation}
with $y^{\e,\nu}_c$ and $M_c$ the dominant contributions to the third and heaviest generation.
Masses and mixing are compatible with the expansion parameters $\epsilon_\e \simeq 0.15$ and $\epsilon_N\sim \epsilon_\nu^3$.
In addition to $y^{\nu}_c$, $Y_\nu$ depends on effective parameters $y^{\nu}_{a,b}$, sourced from the subleading operators in Eq.~\eqref{eq:Wf} and defined explicitly in Eqs.~\eqref{eq:modelparam}--\eqref{eq:paramhierarchy} below.
Due to the flavon VEVs, they correspond to a hierarchy $y^{\nu}_{a} : y^{\nu}_{b} : y^{\nu}_{c} \approx \epsilon_{\nu}^3 : \epsilon_{\nu}^2 : 1$.
The expansion parameter for Dirac neutrinos, $\epsilon_\nu$, is not constrained by phenomenology, but internal consistency of the model requires that it remains perturbative, i.e. $\epsilon_\nu \lesssim 0.5$.
We shall see that numerically viable regions in parameter space correspond to $ \epsilon_\nu \in [0.05,0.5]$.
Note that the large hierarchy between the first two RH neutrinos $N_{1,2}$ and the heaviest one $N_3$ is characteristic of this kind of model \cite{King:2003rf, Ross:2004qn, deMedeirosVarzielas:2005ax, deMedeirosVarzielas:2011wx, Varzielas:2012ss}, where rather different mixing patterns in the quark and lepton sectors are obtained from the same universal Yukawa structures, on the condition that $\braket{\phi_c}$ is dominant in the quark and charged lepton sectors and irrelevant for the neutrino mass matrix.

The structure in Eq.~\eqref{eq:Mgeneric} can written  precisely as
\begin{equation}
\begin{aligned}
    Y_{\e,\nu} &= y_a^{\e,\nu} (\Box_{ab} + \Box_{ba}) + y_b^{\e,\nu} \Box_b + y_c^{\e,\nu} \Box_c , 
    \\ \qquad
    M_N &= M_a (\Box_{ab} + \Box_{ba}) + M_b \Box_b + M_c \Box_c ,
\end{aligned}
\label{eq:YfMNdecomp}
\end{equation}
where $\Box_i = \phi_i \phi^T_i$ and $\Box_{ij} = \phi_i \phi^T_j$ are rank-one matrices,
\begin{equation}
    \Box_{ab} = 
    (\Box_{ba})^T = 
    \begin{pmatrix} 
        0 & 0 & 0 \\
        1 & 1 & -1 \\
        1 & 1 & -1 \\
    \end{pmatrix}, \quad
    \Box_b = 
    \begin{pmatrix} 
        0 & 0 & 0 \\
        0 & 1 & 1 \\
        0 & 1 & 1 \\
    \end{pmatrix}, \quad
    \Box_c = 
    \begin{pmatrix} 
        0 & 0 & 0 \\
        0 & 0 & 0 \\
        0 & 0 & 1 \\
    \end{pmatrix}.
\label{eq:boxes}
\end{equation}
The set of parameters $(y_a^{\e,\nu},y_b^{\e,\nu},y_c^{\e,\nu})$ and $(M_a,M_b,M_c)$ in Eq.~\eqref{eq:YfMNdecomp} are generally complex%
\footnote{%
    The presence of a CP symmetry can constrain them to be real, with CP being broken spontaneously e.g. through the flavon VEVs as in \cite{Bjorkeroth:2015uou}. 
    We don't consider this possibility here.
}
with phases coming either from the VEVs or the coefficients,
\begin{equation}
\begin{aligned}
    y_a^{\e,\nu} &\equiv |y_a^{\e,\nu}| e^{i \gamma_{\e,\nu}}, \qquad &
    y_b^{\e,\nu} &\equiv |y_b^{\e,\nu}| e^{i \delta_{\e,\nu}}, \qquad &
    y_c^{\e,\nu} &\equiv |y_c^{\e,\nu}| ,
\\
    M_a &\equiv |M_a|  e^{i \gamma_N}, \qquad &
    M_b &\equiv |M_b|  e^{i \delta_N}, \qquad &
    M_c &\equiv |M_c| . 
\end{aligned}
\label{eq:modelparam}
\end{equation}
Given that phenomenology depends only on two independent combinations of the phases, we follow \cite{deMedeirosVarzielas:2017sdv} in taking just $\delta$ and $\gamma$ as independent phases (see also Table~\ref{tab:coeff}).
In terms of the fundamental parameters of the superpotential in Eqs.~\eqref{eq:Wf}--\eqref{eq:WN}, they are
\begin{equation}
\begin{aligned}
    |y_a^{\e,\nu}| &= \frac{g_a^{\e,\nu} v_a  v_b \braket{S}}{\sqrt{6} \Lambda^{3}} , &
    |y_b^{\e,\nu}| &= \frac{g_b^{\e,\nu} v_b^2 \braket{\Sigma}}{2 \Lambda^{3}} , &
    |y_c^{\e,\nu}| &= \frac{g_c^{\e,\nu} v_c^2}{\Lambda^2} ,
    \\
    |M_a| &= \frac{g_a^N v_a v_b^2 v_\phi^2}{2\sqrt{3} \Lambda^{4}} , &
    |M_b| &= \frac{g_b^N v_a v_b^2 v_\phi^2}{2 \sqrt{3} \Lambda^{4}} , &
    |M_c| &= \frac{g_c^N v_\phi^2}{\Lambda} .
\end{aligned}
\label{eq:paramhierarchy}
\end{equation}
The superfield $S$ is a gauge singlet, while $\Sigma$ is not \cite{deMedeirosVarzielas:2017sdv} and introduces Clebsch-Gordan (CG) coefficients, although for our purposes here it is sufficient to consider their respective VEVs $\braket{S}$ and $\braket{\Sigma}$ as real numbers, and absorb the different CG contributions to charged leptons and neutrino into $g_b^{e,\nu}$.
The expansion parameters of the model in Eq.~\eqref{eq:Meps} are recovered from the parameters in Eq.~\eqref{eq:YfMNdecomp} as
\begin{equation}
\begin{aligned}
    \epsilon_{\e,\nu} &
    \sim \left| y_a^{\e,\nu}/y_b^{\e,\nu} \right|
    \sim \left| y_b^{\e,\nu}/y_c^{\e,\nu} \right|^{1/2}
    \sim \left| y_a^{\e,\nu}/y_c^{\e,\nu} \right|^{1/3}, 
    \\
    \epsilon_N &
    \sim \left| M_a / M_c \right|^{1/2}
    \sim \left| M_b / M_c \right|^{1/2} .
\end{aligned}
\end{equation}

The lepton asymmetries are obtained in the flavour basis, wherein the charged lepton Yukawa matrix $Y_\e$ and RH neutrino mass matrix $M_N$ are diagonal.
They are diagonalized by unitary matrices, such that
\begin{equation}
\begin{aligned}
    \hat{Y}_\e &= V_{\e L} Y_\e V_{\e R}^\dagger, \\
    \hat{M}_N &= V_N M_N V_N^T ,
\end{aligned}
\end{equation}
where hats ($~\hat{}~$) denote diagonal matrices of positive eigenvalues and, $Y_\e$ being complex symmetric, we have $V_{\e R}^\dagger=V_{\e L}^T$.
In the flavour basis, in the LR phase convention (where the Yukawa couplings are given by $\mathcal{L} \sim \overbar{L} H_d e_R + \overbar{L} H_u \nu_R + \mathrm{h.c.} $), the neutrino Yukawa matrix is given by $\lambda_\nu$, where
\begin{equation}
    \lambda_\nu^* \equiv V_{\e L} Y_\nu V_N^T,
\label{eq:lambdanu}
\end{equation}
where the additional conjugation on $\lambda_\nu$ appears due to the change from the supersymmetry basis to the seesaw basis \cite{Bjorkeroth:2015tsa}.

\section{The UTZ seesaw mechanism}
\label{sec:SeeSaw}

In this section we review the results from \cite{deMedeirosVarzielas:2017sdv} for how the seesaw mechanism operates in the UTZ model, understanding them through a new formulation based on rank-one matrices.
As the Dirac and Majorana matrices are expressed in terms of the same rank-one matrices, the application of the usual seesaw formula,
\begin{equation}
  m_\nu = - v_u^2 Y_\nu M_N^{-1} Y_\nu^T ,
\label{eq:seesaw}
\end{equation}
provides a light neutrino mass matrix $ m_\nu $ which can be expanded in the same fashion, i.e.
\begin{equation}
    m_\nu = m_a (\Box_{ab} + \Box_{ba}) + m_b \Box_b + m_c \Box_c,
\label{eq:YMmatrices3}
\end{equation}
with the $\Box$ matrices defined in Eq.~\eqref{eq:boxes}.
Notably, the UTZ is preserved. 
A detailed discussion of this elegant property can be found in Appendix \ref{app:seesawrank1}.
The parameters $m_{a,b,c}$ entangle the combinations of Dirac and Majorana neutrino couplings as
\begin{equation}
	m_a = -\frac{v_u^2 {y^\nu_a}^2}{M_a} , \qquad
    m_b = m_a \left(2 \frac{y^\nu_b}{y^\nu_a} - \frac{M_b}{M_a} \right), \qquad
    m_c = -\frac{v_u^2 {y^\nu_c}^2}{M_c} .
\label{eq:ma-mc}
\end{equation}
Obtaining the correct neutrino mixing requires $m_c < m_a < m_b$.
In fact, if $m_c < m_a, m_b$, the light neutrino mass matrix in Eq.~\eqref{eq:YMmatrices3} is semi-diagonalized by a tri-bimaximal (TB) rotation (see e.g. \cite{Varzielas:2015aua}). 
Moreover if $m_a < m_b$, the resulting pattern has a Gatto-Sartori-Tonin \cite{Gatto:1968ss} structure which can be fully diagonalized by a rotation of an angle $\theta$ in the $23$ block.
Consequently Eq.~\eqref{eq:YMmatrices3} is compatible with a normal-ordered neutrino spectrum, with
\begin{equation}
    \hat{m}_\nu \equiv \diag\left( m_1 , m_2 , m_3\right) \simeq  
    \diag\left(\frac{|m_c|}{6} , 3  \left|\frac{m_a^2}{m_b}\right| , 2 |m_b| \right) .
\end{equation}
At leading order the full PMNS matrix is given by
\begin{equation}
\label{eq:UPMNS}
    U_\mathrm{PMNS} = 
    \frac{1}{\sqrt{6}}
    \begin{pmatrix}
        2 & \sqrt{2} c_\theta & \sqrt{2} s_\theta \\
        -1 & \sqrt{2} c_\theta - \sqrt{3} s_\theta & \sqrt{3} c_\theta + \sqrt{2} s_\theta \\
        1 & - \sqrt{3} c_\theta - \sqrt{2} s_\theta & \sqrt{2} c_\theta - \sqrt{3} s_\theta
    \end{pmatrix}, \qquad 
    s_\theta = \sqrt{\frac{m_2}{m_3}} \simeq \sqrt{\frac{3}{2}} \left| \frac{m_a}{m_b} \right| .
\end{equation}
In this class of model \cite{King:2003rf, Ross:2004qn, deMedeirosVarzielas:2005ax, deMedeirosVarzielas:2011wx, Varzielas:2012ss} the different mixing patterns in the quark and lepton sectors require a large hierarchy between the first two RH neutrinos and the third, i.e. $ M_{1,2} \ll M_3 $. 
Indeed, given the relations in Eq.~\eqref{eq:ma-mc} and the hierarchy in the Dirac sector (see Eq.~\eqref{eq:Meps}), i.e. $y_a^\nu \sim \epsilon_\nu^3 y_c^\nu$ and $y_b^\nu \sim \epsilon_\nu^2 y_c^\nu$, the requirement that $m_c < m_a < m_b$, implies the following relations for the RH neutrino parameters: $M_a/M_b < \epsilon_\nu$ and $M_a/M_c < \epsilon_\nu^6$. 
Therefore, $N_3$ with $ M_3 \sim M_c $ effectively decouples after seesaw. 
The parameters $M_{a,b}$ are given by the same operator and a moderate hierarchy between them is obtained by the relative size of the coefficients $g_a^N$, $g_b^N$.
For those values of the Dirac neutrino expansion parameter $\epsilon_\nu$ preferred by the model, we thus expect a hierarchical spectrum for the Majorana neutrino masses in which $M_1 < M_2 \ll M_3$.%
\footnote{%
    This different hierarchy in the neutrino Dirac and Majorana matrices can be accommodated in the model through different mediator masses for the Dirac, and RH Majorana mediators, respectively $\Lambda_D$, $\Lambda_N$. 
    From Eqs.~\eqref{eq:Wf} and \eqref{eq:WN}, if $v_b/\Lambda_D \simeq v_a/\Lambda_D \simeq \epsilon_\nu$, requiring $\Lambda_D/\Lambda_N \simeq \epsilon_\nu$ leads to $M_a/M_c < \epsilon_\nu^6$.
}

\section{Leptogenesis}
\label{sec:Leptogenesis}

\subsection{Boltzmann equations}

The generation of a BAU through $N_i$-leptogenesis is a non-equilibrium process which is generally treated by means of Boltzmann equations for the number densities of RH (s)neutrinos, $Y_{N_i}$ and $Y_{\widetilde{N}_i}$ (for an $N_i$ neutrino with mass $M_i$), and leptons, $Y_{L_\alpha}$. 
It is useful to consider the quantities $Y_{\Delta_\alpha} = Y_{B}/3 - Y_{L_\alpha}$ rather than $Y_{L_\alpha}$, since $\Delta_\fli = B/3 - L_\alpha$ is conserved by sphalerons and other SM interactions. 
$L_\alpha$ and $\Delta_\alpha$ asymmetries are related by a flavour coupling matrix $A$, i.e. $Y_{L_\fli} = \sum_{\fli^\prime} A_{\fli\fli'} Y_{\Delta_{\fli}}$.
The form of $A$ depends on which interactions are in thermal equilibrium during leptogenesis; it is defined explicitly in Appendix \ref{app:boltzmann}.
The produced lepton asymmetries are partially converted into a baryon asymmetry $Y_B$ by the sphalerons, given in the MSSM by
\begin{equation}
    Y_B = \frac{10}{31} \sum_{\fli} Y_{\Delta_\fli},
\end{equation}
with $Y_{\Delta_\fli}$ computed at a temperature $T \ll M_i$, where the densities $Y_{N_i} $, $Y_{\widetilde{N}_i}$ are effectively zero.
In the fully flavoured regime, $ M_i \ll 10^9 (1+\tan^2\beta)$ GeV, all lepton flavours are to be treated separately, i.e. $ \fli = e, \mu, \tau $.
In the two-flavour regime, $ 10^9 (1+\tan^2\beta) \mathrm{~GeV} \ll M_i \ll 10^{12} (1+\tan^2\beta) $ GeV, only the interaction mediated by the $\tau$ Yukawa coupling is in equilibrium and the asymmetries in the $e$ and $\mu$ flavours can be treated with a combined density $Y_{\Delta_{e\mu}} = Y_{\Delta_e+\Delta_\mu}$.

In the MSSM, with hierarchical RH neutrinos only $N_1$ and $N_2$ participate in the leptogenesis process and we can neglect the contribution from $N_3$. We have three different scenarios, depending on $M_1$ and $M_2$. Assuming first that both $M_1, M_2 \ll 10^9 ( 1 + \tan ^2 \beta)$ GeV, the three charged-lepton states are active in the plasma during leptogenesis and the Boltzmann equations take the form \cite{Antusch:2006cw}
\begin{equation}
\begin{aligned}
    \frac{d Y_{N_i}}{dz} &= 
        - 2 D \left(Y_{N_i}-Y_{N_i}^\mathrm{eq}\right) , \\
    \frac{d Y_{\widetilde{N}_{i}}}{dz} &= 
        - 2 D\left(Y_{\widetilde{N}_{i}}-Y_{\widetilde{N}_{i}}^\mathrm{eq}\right) , \\
    \frac{d Y_{\Delta_\fli}}{dz} &= 
         2 \cpasym{\fli}{i} D\left(Y_{N_i}-Y_{N_i}^\mathrm{eq}\right) + 2 \cpasyms{\fli}{i} D \left(Y_{\widetilde{N}_{i}}-Y_{\widetilde{N}_{i}}^\mathrm{eq}\right)
        + \frac{\decK{\fli}{i}}{\decK{}{i}} W \sum_\fli A_\mathrm{\fli \fli^\prime} Y_{\Delta_{\fli^\prime}} ,
\end{aligned}
\label{eq:Boltz}
\end{equation}
where $z=M_i/T$ and $Y_{N_{i}}^\mathrm{eq}$, $Y_{\widetilde{N}_i}^\mathrm{eq}$ are the equilibrium densities of (s)neutrinos $ N_i $, $ \widetilde{N}_i $, respectively. 
In this case, the flavour index $\alpha$ runs over the three lepton flavours, $\alpha = e, \mu, \tau$, and the asymmetries $Y_{\Delta_\fli}$ are stored separately in the different flavours.
The factors $D$ and $W$ govern the decay and washout behaviour, respectively, and contain information about decays, inverse decays, and scattering processes. 
\cite{Davidson:2002qv, Giudice:2003jh, Buchmuller:2004nz, Abada:2006fw}.
The expressions used in our calculation are collected in Appendix \ref{app:boltzmann}, where we follow in particular the notation and method of \cite{Antusch:2006cw}. The decay factors $\decK{\fli}{i}$ and CP asymmetries $\cpasym{\fli}{i}$, arising from the interference between tree-level and loop diagrams of the RH neutrino decay, are determined by the flavour parameters of the model, and are explored in the next subsection.

In the case $ M_1 < 10^9 (1+\tan^2\beta)~\mathrm{GeV} < M_2 $, only the tau Yukawa coupling is in equilibrium during $N_2$ decays. We thus have two lepton flavours in the process, $\alpha = \tau, (e+\mu)$. 
In this first step, two asymmetries are generated, $Y_{\Delta_{\tau}} $ and $Y_{\Delta_{e\mu}}$, following  Eq.~\eqref{eq:Boltz}. 
However, before the decay of $N_1$, the muon Yukawa coupling reaches equilibrium and $Y_{\Delta_{e\mu}}$ is projected on the $e$ and $\mu$ flavours, proportionally to  $\decK{e}{2}$ and $\decK{\mu}{2}$, respectively. 
We then use these values as initial conditions in $N_1$ decays, using again Eq.~\eqref{eq:Boltz} with $\alpha = e, \mu \tau$.

Finally, we can have both $N_1$ and $N_2$ in the two-flavour regime, $10^9 (1+\tan^2\beta) \mathrm{~GeV} \ll M_1, M_2$. 
An asymmetry is generated from $N_2$ decays in the $(e+\mu)_2$ flavour, i.e. in the combination of $e$ and $\mu$ that couples to $N_2$. 
This combination maintains the coherence in the plasma between a decay and a subsequent inverse decay. 
Then, when $T \sim M_1$, the couplings of $N_1$ select a different combination of of $e$ and $\mu$ in the direction of the $N_1$ Yukawa coupling, $(e + \mu)_1$. 
This implies that only the component of the $(e+\mu)_2$ asymmetry in the $(e+\mu)_1$ direction can be washed-out by $N_1$ inverse decays, while the rest, orthogonal to $(e+\mu)_1$, remains untouched by $N_1$.

Note that in all numerical calculations below, we use the instantaneous approximation to describe the transition between the two-flavour and three-flavour regime.  
A more rigorous description of the transition between two different flavoured scenarios would require the use of density matrix equations, as noted in \cite{Barbieri:1999ma,Abada:2006fw} and described in detail in \cite{Blanchet:2011xq}.

\subsection{Decay factors and CP asymmetries}
\label{sec:Param}

The lepton asymmetry in each flavour is governed by two sets of parameters which can be computed within a given neutrino model: the decay factors $\decK{\fli}{i}$ and CP asymmetries $\cpasym{\fli}{i}$, for a neutrino $N_{i}$ decaying into a Higgs $H_u$ and lepton doublet $L_\fli$ (or their conjugates). 
The Majorana nature of the RH neutrino masses implies the decays $N_{i} \rightarrow L_\fli H_u$ and $N_{i} \rightarrow \overbar{L}_\fli H_u^{*}$ violate lepton number by one unit ($\Delta L = 1$).
The decay factors are defined as
\begin{equation}
    \decK{\fli}{i} = \frac{\Gamma(N_{i}\rightarrow{L}_\fli H_u) 
        + \Gamma(N_{i} \to \overbar{L}_\fli H_u^*)}{{\mathcal{H}}(M_{i})},
    \qquad 
    \decK{}{i}=\sum_{\fli} \decK{\fli}{i} ,
\end{equation}
where $\mathcal{H}(T)$ is the Hubble parameter at the temperature $T$, and $\mathcal{H}(M_i) \simeq 1.66 \sqrt{g_*} M_i^2/M_\mathrm{Pl}$. 
The CP asymmetries are defined as 
\begin{equation}
    \cpasym{\fli}{i} = \frac{\Gamma(N_{i} \rightarrow L_\fli H_u) - \Gamma(N_{i} \rightarrow \overbar{L}_\fli H_u^{*})}{\Gamma(N_{i} \rightarrow L_\fli H_u) + \Gamma(N_{i} \rightarrow \overbar{L}_\fli H_u^{*})} .
\end{equation}
The decay factors are dominated by the single tree-level diagram, while the CP asymmetries arise only at one-loop level from the self-energy plus vertex diagrams.
In the two-flavour regime, $\decK{e\mu}{i} = \decK{e}{i} + \decK{\mu}{i} $, with the corresponding decay asymmetry $\cpasym{e\mu}{i} = \cpasym{e}{i} + \cpasym{\mu}{i}$.

Explicitly in terms of the neutrino Yukawa matrix in the flavour basis, $\lambda_\nu$, the decay factors are given by
\begin{equation}
    \decK{\fli}{i} = \frac{v_u^2}{m_\ast M_i} (\lambda_\nu^\dagger)_{i \fli}(\lambda_\nu)_{\fli i},
\end{equation}
where $m_\ast \simeq (1.58 \times 10^{-3} \mathrm{~eV}) \sin^2 \beta$. 
For $N_{i}$ decay, the relative phase between the tree diagram and the loop diagram with an intermediate $N_{j}$ will be the phase of $(\lambda_\nu^\dagger \lambda_\nu)_{ij}$. 
Then the CP-asymmetries for the two lightest RH-neutrinos is expressed as
\begin{equation}
    \cpasym{\fli}{i} = \frac{1}{8\pi}  \sum_{j \neq i} \frac{
    \mathrm{Im}[(\lambda_\nu^\dagger)_{i\fli} (\lambda_\nu^\dagger \lambda_\nu)_{ij} (\lambda_\nu)_{\fli j}]}
    {(\lambda_\nu^\dagger \lambda_\nu)_{ii}}
    g\bigg(\frac{M_j^2}{M_i^2}\bigg) .
\label{eq:epsilon1}
\end{equation}
where $g(x)$ is a loop function given by the sum of the vertex and the self energy contributions \cite{DiBari:2005st,Antusch:2006cw}; in the MSSM,
\begin{equation}
    g(x) = \sqrt{x}\left[\frac{2}{1-x}-\log\left(\frac{1+x}{x}\right)\right] .
\end{equation}

An exploration of the CP asymmetries and decay factors -- responsible for the production and washout of a lepton asymmetry, respectively -- provides some insight into how leptogenesis proceeds in this model.
The decay factors appear in the arguments of exponential damping terms, and a large $ \decK{\fli}{i} $ is associated with strong washout.
As it is inversely proportional to the RH neutrino mass, in the ``vanilla'' picture of flavour-independent $N_1$ leptogenesis, this yields a lower bound on the $N_1$ mass, $M_1 \gtrsim 10^9$ GeV \cite{Davidson:2002qv}.
When considering asymmetry generation from next-to-lightest RH neutrinos ($N_2$ leptogenesis), typically a crucial requirement is that $\decK{\fli}{1} \lesssim 1$ in some lepton flavour, to not completely wash out a previously generated asymmetry from $N_2$ decays \cite{Buchmuller:2004nz}.
This depends in particular on the Yukawa structures that give $\lambda_\nu$; $N_2$ leptogenesis and its compatibility with low-scale neutrino phenomenology has been studied in \cite{DiBari:2010ux, DiBari:2013qja, DiBari:2014eya, DiBari:2014eqa}.

As we are considering the case in which $M_{1}\ll M_{3}$ then we can also neglect the $i=3$ contribution to $ \cpasym{\fli}{1}$. 
In Appendix \ref{app:LpGnparam} we show that $\lambda_\nu$ (in the flavour basis) maintains the hierarchical structure suggested by the model, and a rough estimate for the leptogenesis parameters gives
\begin{equation}
\begin{aligned}
    \decK{\fli}{1} &\sim \left|\frac{M_b M_c}{M_a^2}\right| \epsilon_\nu^6
        \pmatr{ \epsilon_\nu^2\\1\\1 } , \quad & 
    \cpasym{\fli}{1} &\sim \frac{3}{8\pi} \left|\frac{M_a}{M_b}\right|^2 \epsilon_\nu^4 
        \pmatr{ \epsilon_\nu^2\\1\\1 } , \\
    \decK{\fli}{2} &\sim \left|\frac{M_c}{M_b}\right| \epsilon_\nu^4 
        \pmatr{ \epsilon_\nu^2\\1\\1} , &
    \cpasym{\fli}{2} &\sim \frac{3}{8\pi} \left|\frac{M_b}{M_c}\right| 
        \pmatr{ \epsilon_\nu^4\\\epsilon_\nu^2\\1 } 
        + \frac{1}{4\pi} \left| \frac{M_a}{M_b} \right|^2 \epsilon_\nu^6 \pmatr{ \epsilon_\nu^2 \\ 1 \\ 1}.
\end{aligned}
\label{eq:lepto3gen}
\end{equation}

From this we can make some \emph{a priori} considerations:
(i) due to the UTZ in the electron coupling to $N_1$, lepton asymmetries from $N_1$ decays are dominated by the $\mu$ and $\tau$ flavours, while $\cpasym{e}{1}$ is generally too small to contribute significantly to asymmetry production, 
(ii) we similarly expect a strong washout in the $\mu$ and $\tau$ flavours for both $N_1$ and $N_2$ leptogenesis, and comparatively weak washout for the electron, and
(iii) despite the large hierarchy between $M_b$ and $M_c$, the $\cpasym{\fli}{2}$ are typically dominated by the first term, which generates non-negligible asymmetry only if $M_b$ and $M_c$ are not too separated.

\section{Analysis and results}
\label{sec:Analysis}

\subsection{Numerical results}
\label{sec:numericalresults}

In this section we present the numerical solutions to the fully flavoured Boltzmann equations in the MSSM as given in Section~\ref{sec:Leptogenesis}, following a similar procedure to the one already adopted in \cite{Bjorkeroth:2016lzs}. 
The analysis has been performed under the assumption that the spectrum of the heavy neutrinos in the model is hierarchical, $M_1 < M_2 \ll M_3$. 
Within this framework, 
\begin{itemize}
    \item a possible asymmetry generated by the heaviest RH neutrino $N_{3}$ is always washed out and assumed to be negligible,
    \item the generation of the asymmetry and the washout from decays and inverse decays of the $N_1$ neutrinos starts only after the end of the analogous processes from the $N_2$.
    The two lightest RH neutrinos do not interfere with each other, such that the generation of the asymmetry from $N_1$ decays and from $N_2$ decays proceed independently.
\end{itemize} 
Consequently, the Boltzmann equations in Eq.~\eqref{eq:Boltz} are solved twice for each point in the model parameter space.
In the first step, we solve for $Y_{\Delta_\fli}$ arising from $N_{i=2}$ decays, assuming thermal initial conditions (zero neutrino and asymmetry densities).
The solutions for $Y_{\Delta_\fli}$ are then used as initial conditions for the $N_{i=1}$ calculation. 
The final asymmetry is obtained from the sum over $Y_{\Delta_\fli}$ after $N_1$ leptogenesis.

The input parameters are comprised of those not already fixed by the fit to low-scale neutrino phenomenology.
In particular, we use the fit to quark and lepton masses and mixing for our flavour model performed in \cite{deMedeirosVarzielas:2017sdv}, with relevant best fit values for the lepton sector given in Table~\ref{tab:coeff}.
We fix $\tan \beta = 10$ in this analysis.
Note also that, as the model does not determine the absolute mass scale for fermions, the fit only provides estimates for the parameters in Eq.~\eqref{eq:modelparam} up to an overall scale, which is set by the third generation, i.e. by $M_c$ and $y_c^{\e,\nu}$.
With $\tan \beta$ fixed, we can infer the charged lepton scale $y^\e_c$, while the neutrino scales $y^\nu_c$ and $M_c$ remain unfixed.

\begin{table}[ht]
\centering
\begin{tabular}{ccccc}
\toprule
    \multicolumn{2}{c}{{Neutrinos}} && \multicolumn{2}{c}{{Charged leptons}} \\
\cmidrule{1-2} \cmidrule{4-5}
    $m_a/$meV   & 8.95 && $y_a^\e$      & $3.01 \times 10^{-4}$ \\ 
    $m_b/$meV   & 24.6 && $y_b^\e$      & $3.90 \times 10^{-3}$ \\
    $m_c/$meV   & 2.26 && $y_c^\e$      & $7.16 \times 10^{-2}$ \\
    $\gamma_m$  & 2.51 && $\gamma_\e$   & $0.13$ \\ 
    $\delta_m$  & 1.26 && $\delta_\e$   & $-1.31$ \\
\bottomrule
\end{tabular}
\caption{Fitted values for the low-scale model parameters, extracted from the computation in \cite{deMedeirosVarzielas:2017sdv}.}
\label{tab:coeff}
\end{table}

The fit fixes the values of the neutrino mass parameters $m_{a,b,c}$ and charged lepton parameters $y_{a,b,c}^\e$.
The seesaw relation in Eq.~\eqref{eq:ma-mc} entangles the three Dirac and three Majorana neutrino couplings ($y^\nu_{a,b,c}$ and $M_{a,b,c}$, respectively), constrained only by the three fitted values of $m_{a,b,c}$, leaving three (real) free parameters which enter into the leptogenesis analysis.
The phases of the couplings are similarly related: $Y_\nu$ and $M_N$ each contain two independent phases ($\gamma_\nu, \delta_\nu$ and $\gamma_N, \delta_N$, respectively); combinations of these yield the two fitted independent phases of $m_\nu$. 

\begin{figure}
\centering
    \begin{subfigure}[b]{0.50\textwidth}
       \includegraphics[width=\textwidth]{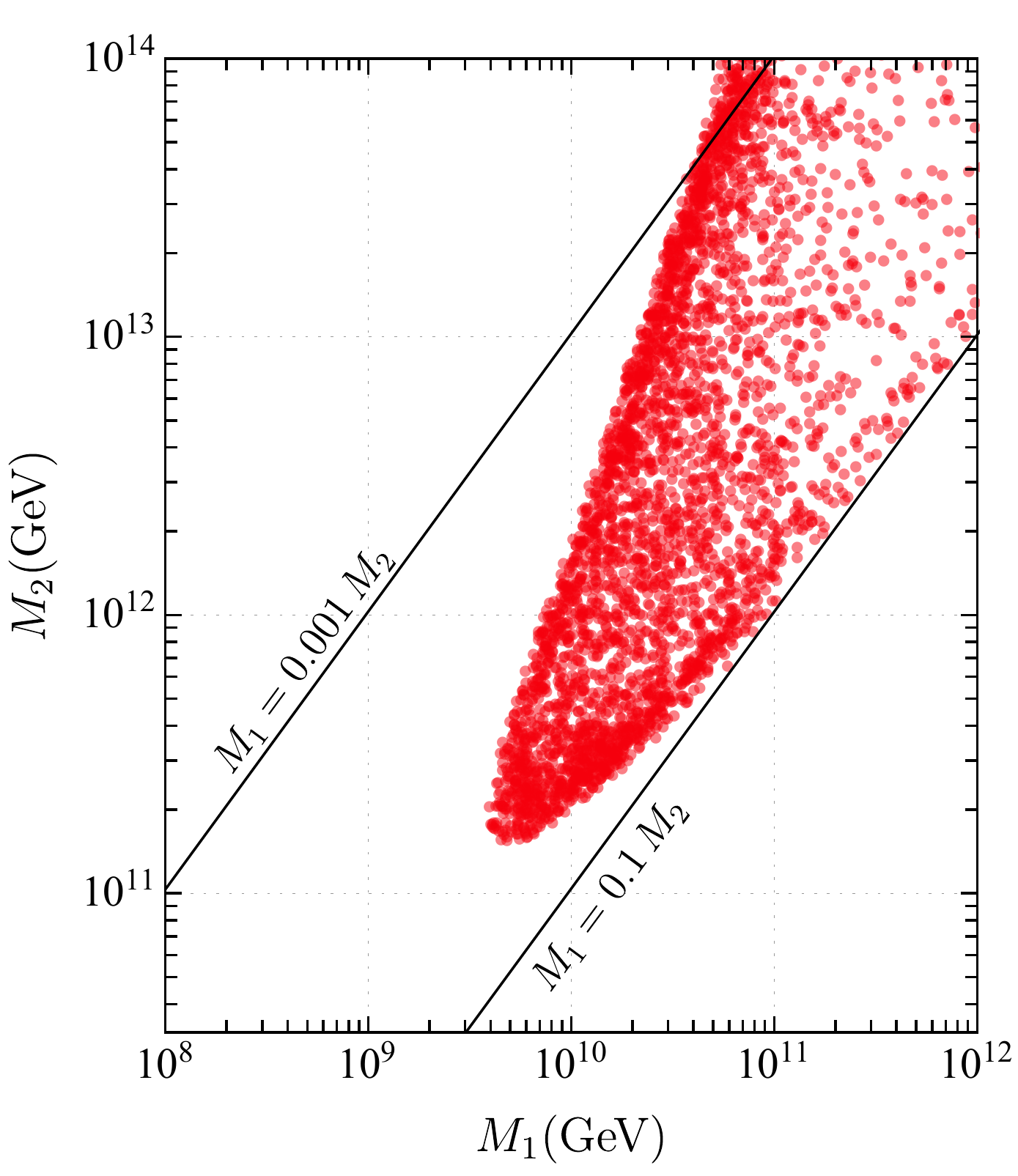}
    \caption{}
    \label{fig:M1M2a}
    \end{subfigure}%
    \begin{subfigure}[b]{0.50\textwidth}
        \includegraphics[width=\textwidth]{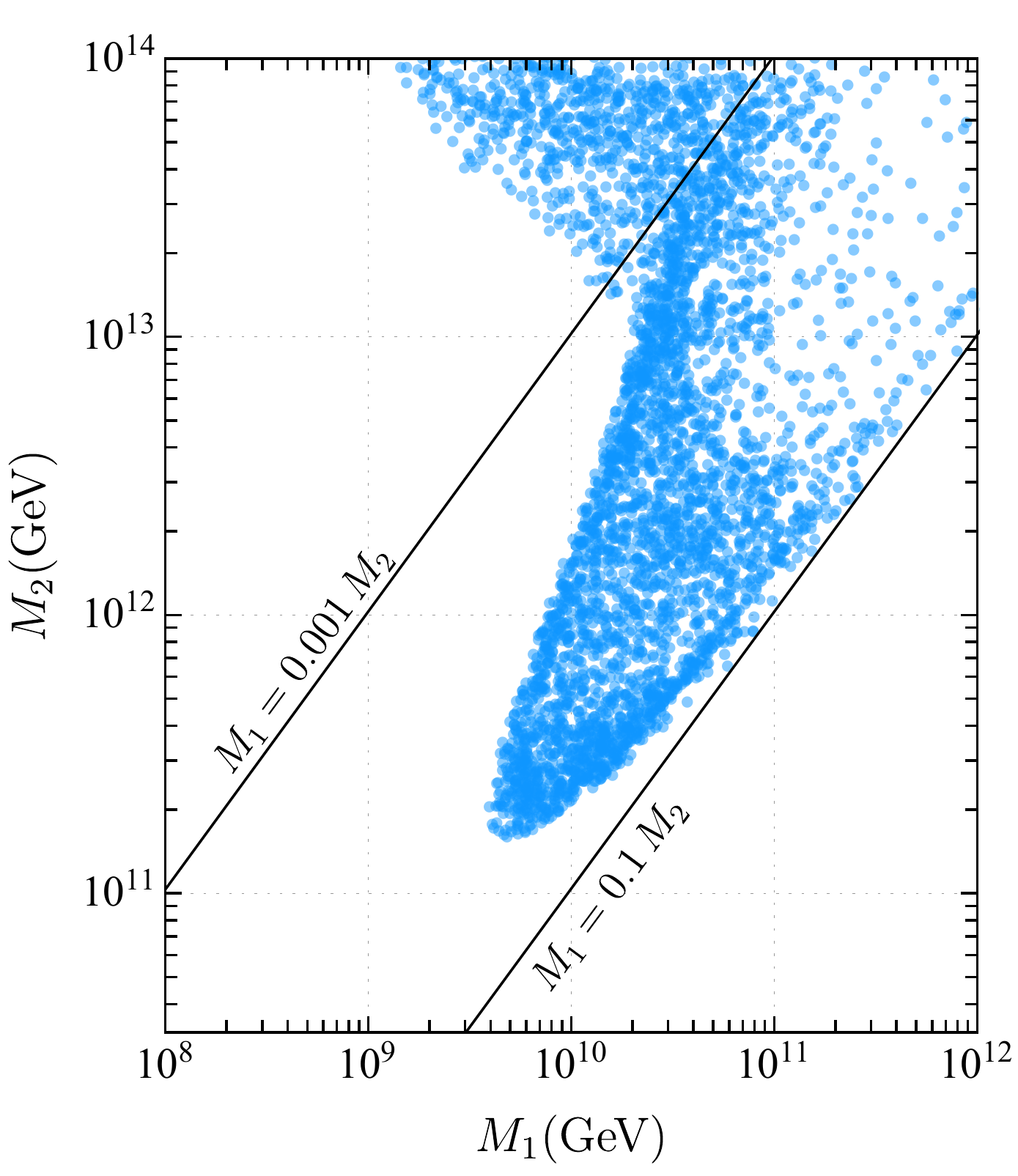}
    \caption{}
    \label{fig:M1M2b}
    \end{subfigure}
\caption{Allowed values of RH neutrino eigenvalues $M_{1,2}$ giving $Y_B$ within $20\%$ of the observed value. The $N_3$ mass is fixed to be $M_3 = 5 \times 10^{14}$ GeV. Red points assume the only contributions are from $N_1$ decays, while blue points take into account both $N_1$ and $N_2$ decays.}
\label{fig:M1M2}
\end{figure}

\begin{figure}
\centering
    \begin{subfigure}[b]{0.50\textwidth}
       \includegraphics[width=\textwidth]{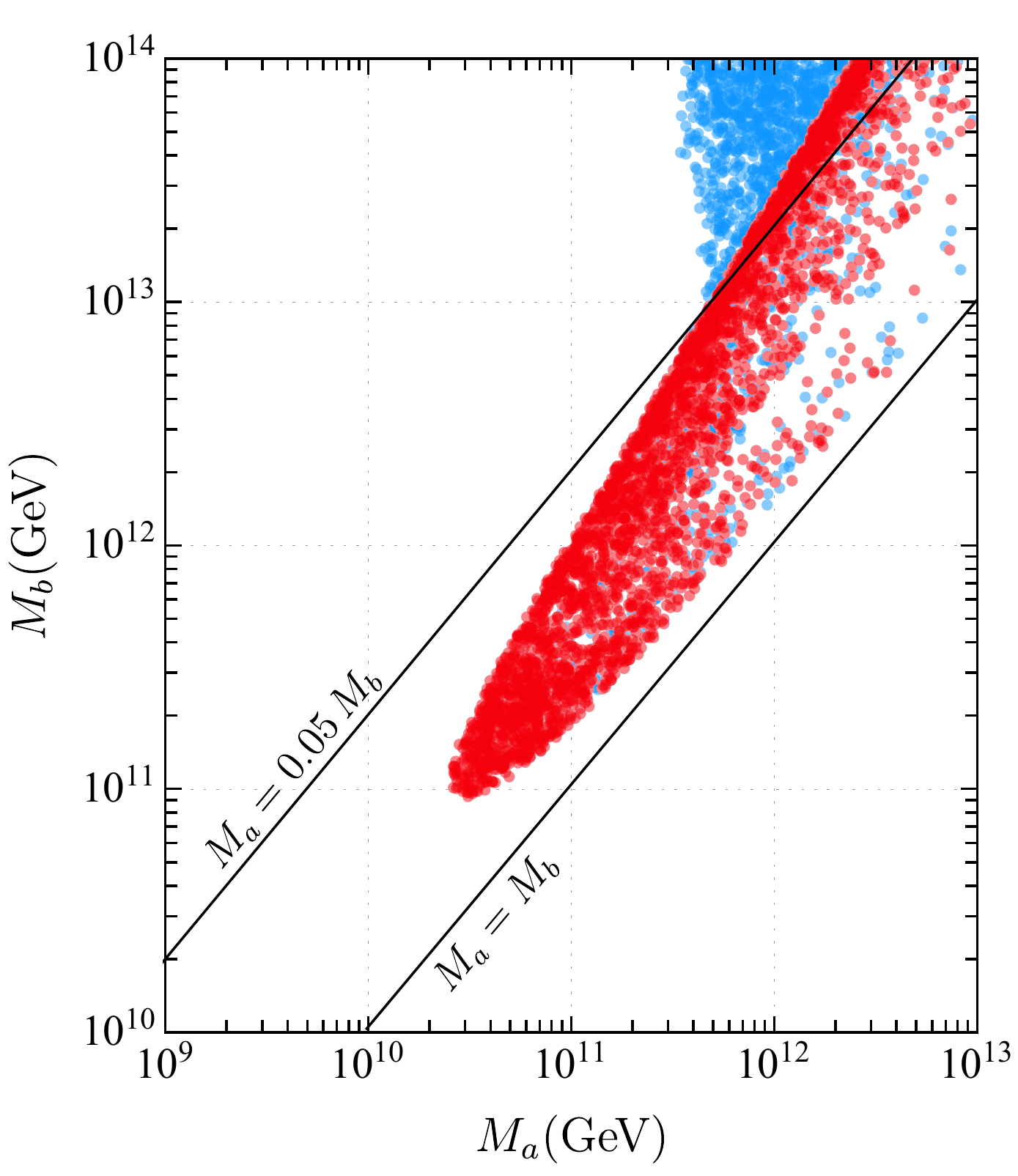}
    \caption{}
    \label{fig:MaMb}
    \end{subfigure}%
    \begin{subfigure}[b]{0.49\textwidth}
        \includegraphics[width=\textwidth]{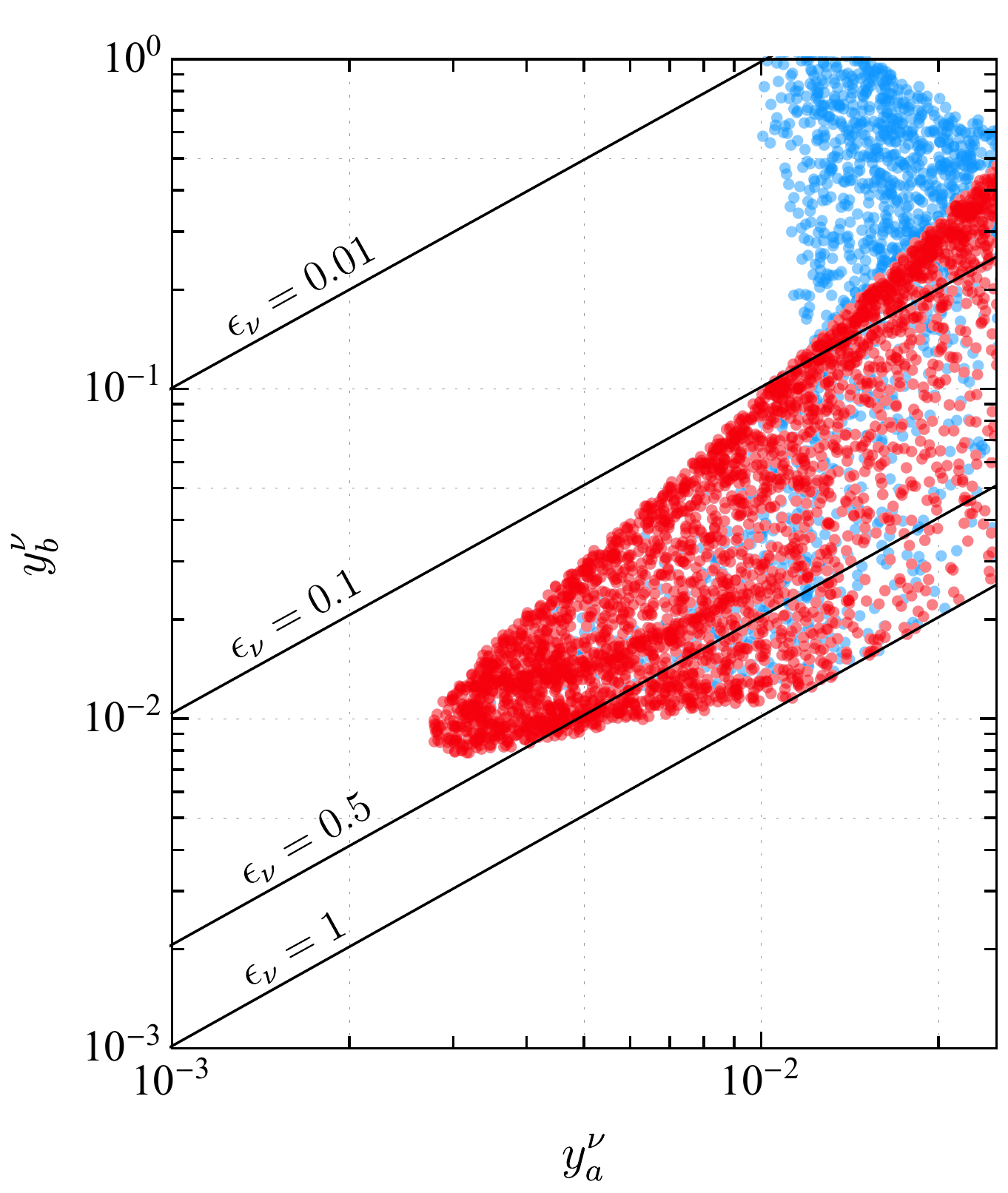}
    \caption{}
    \label{fig:yayb}
    \end{subfigure}
\caption{Allowed values of RH input mass parameters $M_{a,b}$, Dirac neutrino couplings $y_{a,b}^\nu$, and RH neutrino masses $M_{1,2}$ giving ${Y}_B$ within $20\%$ of the observed value. The colours correspond to those Figure~\ref{fig:M1M2}.}
\label{fig:MaMbyayb}
\end{figure}

For this analysis, we choose the sets $M_{a,b,c}$ and $\gamma_N, \delta_N$ as the inputs, scanning over the ranges $|M_{a,b}| \in [10^7, 10^{14}]$ GeV with fixed $ |M_c| = 5 \times 10^{14}$ GeV, and $\gamma_N, \delta_N \in [-\pi,\pi]$.
For each point, we solve the $N_1$ and $N_2$ Boltzmann equations for $z \in [0,\infty]$. 
We stress that, as the parameters of $Y^\e$ and $m_\nu$ are fixed by the fit, each point automatically satisfies current experimental bounds on lepton masses and mixing. 
The results are shown in Figures~\ref{fig:M1M2} and \ref{fig:MaMbyayb}.

Figure~\ref{fig:M1M2} shows the regions that reproduce  the experimental value of $Y_B$ to within $20\%$, in terms of the RH neutrino mass eigenvalues $M_{1,2}$, with $M_3 \simeq 5 \times 10^{14}$ GeV.
In Figure~\ref{fig:M1M2a} we see the successful leptogenesis regions taking into account only $N_1$ decays, while in Figure~\ref{fig:M1M2b} we consider both $N_1$ and $N_2$ decays. 
The comparison between plots allows us to conclude that, over most of the parameter space of the model, the BAU is consistent with leptogenesis proceeding entirely from $N_1$ decays, assuming thermal initial conditions.
In other words, the asymmetries generated by $N_2$ decays are efficiently washed out in all flavours. 
In the $N_1$ case, the dominant contributions to the viable regions are from the $\mu$ and $\tau$ asymmetries, while the electron asymmetry is completely negligible. 
This agrees well with the expectations from the analytical approximations in Section~\ref{sec:Leptogenesis}.
Nevertheless, in  Figure~\ref{fig:M1M2b} we find a small region where the $N_2$ contribution to the BAU dominates. 
We can see that this scenario requires a small splitting between the heavy RH neutrinos, $M_2/M_3 \gtrsim 0.1$. 
The $N_2$ case is discussed further in Section~\ref{sec:N2lept}, where we show that this region of parameter space is not natural in the UTZ model.

Figure~\ref{fig:MaMb} displays the regions corresponding to $Y_B$ within 20\% of the observed value, in terms of the RH neutrino mass parameters $M_{a,b}$, with $M_c = 5\times 10^{14}$ GeV.
Most of the points correspond to $N_1$ leptogenesis (red points) with $ M_a/M_b \in [0.05, 1]$. 
Figure~\ref{fig:yayb} shows the corresponding regions in terms of the neutrino Dirac couplings $y_{a,b}^\nu$, and the expansion parameter $\epsilon_\nu = y_a^\nu / y_b^\nu$. 
In $N_1$-dominated leptogenesis, we have $y_a^\nu \in [0.003,0.03]$, $y_b^\nu \in [0.008,0.5]$ and $\epsilon_\nu \in [0.05,1]$.

Recalling that $M_1 \sim M_a^2/M_b$ and $M_2 \sim M_b$, the eigenvalues $M_{1,2}$ display a bigger hierarchy when compared to $M_{a,b}$ and these points satisfy $ M_1/M_2 \in [0.002, 0.1]$. 
Therefore, we conclude that the correct BAU is found for RH neutrino masses above $M_1 \gtrsim 4 \times 10^9$ GeV and $M_2 \gtrsim 2 \times 10^{11}$ GeV.
In this regime it is relevant to discuss the issues related to the potential overproduction of gravitinos \cite{Kawasaki:2008qe}. 
There are several ways around it \cite{Jedamzik:2006xz, Pradler:2006qh}, one of which is to keep the reheating temperature low and to produce the RH neutrinos non-thermally (e.g. produced in decays of the inflaton). 
Nonetheless, even for thermal production scenarios, if the gravitino is unstable with mass $ m_{3/2} \gtrsim 10$~TeV, these relatively high reheating temperatures around $10^9$ or $10^{10}$ GeV remain borderline viable.

Finally, it is interesting to analyse the restrictions that a requirement of successful leptogenesis set on the flavour model. 
As we have seen in Figures~\ref{fig:M1M2} and~\ref{fig:MaMbyayb}, the best possibility, if we demand a relatively low reheating temperature, would correspond to RH eigenvalues $M_1 \simeq 4 \times 10^9$ GeV and $M_2 \simeq 2 \times 10^{11}$ GeV, with $M_3 \geq 10^{14}$ GeV. 
In terms of the model parameters these points correspond roughly to $M_a \simeq 3 \times 10^{10}$ GeV, $M_b \simeq 10^{11}$ GeV, $y_a^\nu \simeq 0.003$ and $y_b^\nu \simeq 0.009$. 
We emphasize again that independent information on the neutrino Yukawa couplings and RH neutrino masses is not available from oscillation experiments, but when the BAU is accounted for we can obtain several unknown parameters. 
With the above values we obtain the expansion parameter $\epsilon_\nu \simeq 0.3$. 
The heaviest RH neutrino is then $M_c \times g^\nu_b/g^\nu_c \simeq 2 \sqrt{3}  M_b/\epsilon_\nu^6 \simeq 5 \times 10^{14}$ GeV. 
However, these restrictions depend strongly on the details of the flavour model and may change with small variations \cite{deMedeirosVarzielas:2018vab}. 
If supersymmetry is found in the neighbourhood of the electroweak scale, we would obtain additional information on the flavour symmetry that could help restrict these possibilities \cite{Das:2016czs, Lopez-Ibanez:2017xxw, deMedeirosVarzielas:2018vab, Lopez-Ibanez:2019rgb}.

\subsection{\texorpdfstring{\boldmath{$N_2$}}{N2} leptogenesis and comparison with other models}
\label{sec:N2lept}

As we have seen in the comparison of Figures~\ref{fig:M1M2a} and \ref{fig:M1M2b}, there is a small region where the BAU is generated mainly by $N_2$ leptogenesis. 
This region corresponds to $M_1 \leq 0.002 M_2$, with $M_2 \geq 10^{13}$ GeV and $M_3 = 5 \times 10^{14}$ GeV.
These points correspond to relationships between model parameters, $M_a < 0.002 M_b$ and $0.01<\epsilon_\nu<0.1$.

Here, the mechanism of asymmetry generation and washout is slightly more involved \cite{Barbieri:1999ma,Abada:2006fw,Nardi:2006fx,Antusch:2010ms}.
At temperatures $ T \sim M_2 \sim 10^{13} $ GeV, a comparatively large asymmetry is generated in each of the two active lepton flavours $\fli = (e\mu), \tau$ from $N_2$ decays.
These serve as initial conditions of the subsequent $N_1$ system, which occurs at much lower temperatures $ T \sim M_1 \sim 10^9$ GeV.
This lies in the fully flavoured regime, wherein the active flavours are $\fli = e, \mu, \tau$.
The asymmetry $Y_{\Delta_{e\mu}}$, initially generated in the combined $e\mu$ flavour, is split into the $e$ and $\mu$ flavours proportionally to $\decK{e}{2} \propto |\lambda_{e 2}|^2$ and $\decK{\mu}{2} \propto |\lambda_{\mu 2}|^2$, respectively \cite{Abada:2006fw, Nardi:2006fx, Antusch:2010ms}.%
\footnote{%
    In principle, we should also include the so-called phantom terms \cite{Antusch:2010ms}.
    However, for $\delta = e, \mu$ and using 
    $ p_\delta = \big( 
            \varepsilon_{N_2}^\delta - \varepsilon_{N_2}^{(e+\mu)} K^\delta_{N_2}/K^{(e+\mu)}_{N_2} 
        \big) N_{N_2}^\mathrm{in} $, 
    it is straightforward to check from Eq.~\eqref{eq:lepto3gen} that, even assuming an initial $N_2$ abundance $N_{N_2}^\mathrm{in}$, these terms are always subdominant in our scenario, at least by $\epsilon_\nu^2$, with respect to $\epsilon_{N_2}^{(e + \mu)} K^\delta_{N_2}/K^{(e+\mu)}_{N_2}$. 
    They can therefore be safely neglected in the model considered.
}
Assuming the number density of $e$ and $\mu$ asymmetries are equal at the moment where $\mu$ couplings reach equilibrium, the initial conditions for the $N_1$ decays are thus $ Y_{\Delta_e} \simeq x^2 Y_{\Delta_{e\mu}} $ and $Y_{\Delta_\mu} \simeq (1- x^2) Y_{\Delta_{e\mu}} $, where $ x = |y^\nu_a / y^\nu_b - M_a/M_b|$.

As the CP asymmetries $\cpasym{\fli}{1}$ are sensitive to the ratio $M_1/M_2 \ll 1$, no significant additional contribution to the BAU is generated by $N_1$ decays in this regime.
However, if a large asymmetry is generated by $N_2$ decays, even a small portion stored in the electron flavour can survive washout and reproduce the observed asymmetry.
To understand this, we make two observations: 1) the decay factors $\decK{\fli}{1}$ are approximately proportional to $1/M_1$, and 2) in each lepton flavour, they go like $ (\decK{e}{1}, \decK{\mu}{1}, \decK{\tau}{1}) \sim (\epsilon_\nu^2, 1, 1) $. 
In other words, the flavour structure of the model implies the washout in the electron flavour is generally weaker than other flavours.
Indeed, we observe that the $Y_{\Delta_\mu}$ and $Y_{\Delta_\tau}$ asymmetries are efficiently washed out, while some portion of $Y_{\Delta_e}$ remains.

So, as we can see, $N_2$ leptogenesis is possible (in part) due to the texture zero, which is enforced by symmetry. 
However, from the perspective of the UTZ model based on $\Delta(27)$, described above, this $N_2$-dominated scenario is not ``natural'', while $N_1$ leptogenesis is still viable and natural in large parts of the parameter space.
This unnaturalness is a direct consequence of the structure of the neutrino matrices (see Eq.~\eqref{eq:YfMNdecomp}) and can be understood by looking at Eqs.~\eqref{eq:ma-mc}--\eqref{eq:UPMNS}. 
Using Eq.~\eqref{eq:UPMNS} with the measured value for $\sin \theta_{13} \simeq 0.15$, we obtain 
\begin{equation}
    2 \frac{y_b^\nu}{y_a^\nu}-\frac{M_b}{M_a} \simeq 8.2
\label{eq:sinth13}
\end{equation}
Barring accidental cancellations, this expression fixes $|M_2/M_1| = |M_b^2/M_a^2| \approx 8$.
As a consequence, the structure of Yukawa matrices enhances the leptogenesis effects from $N_1$, proportional to $M_1/M_2$, and suppresses $N_2$ effects, proportional to $M_2/M_3$ (see Eq.~\eqref{eq:lepto3gen}).

$N_2$ leptogenesis can be important in situations where $M_2/M_1 \gg 1$, as seen in Figure~\ref{fig:M1M2}, but this requires a strong cancellation of several orders of magnitude in Eq.~\eqref{eq:sinth13}. 
Moreover, the structure of the neutrino Yukawa matrices in the UTZ model is not hierarchical in this region, as we have $0.1 \leq y_b^\nu \lesssim 1$ while $y_c^\nu \simeq 0.1$.
These values are not natural to the UTZ model, where most of the flavon VEVs are required to be much smaller than 1. 
In conclusion, $N_2$ leptogenesis is possible, but disfavoured. 

This can be compared with the situation in typical $SO(10)$ models \cite{Vives:2005ra,DiBari:2014eya,DiBari:2015oca} and, in general, in models of sequential dominance (SD) \cite{Bjorkeroth:2016lzs}. 
In these models, the three LH neutrino mass scales are each determined independently by a single RH neutrino. 
Schematically, 
SD in the limit of $M_1 < M_2 \ll M_3$ gives
\begin{equation}
    m_1 \propto \frac{{y_3^\nu}^2 v^2_u}{M_3}, \qquad 
    m_2 \propto \frac{{y_1^\nu}^2 v^2_u}{M_1}, \qquad 
    m_3 \propto \frac{{y_2^\nu}^2 v^2_u}{M_2},
\label{eq:mieqdom}
\end{equation}
with a strong hierarchy of $ m_1 < m_2 < m_3$, and where
$y_3^\nu \propto m_t$, $y_2^\nu \propto m_c$ and $y_1^\nu \propto m_u$ (and thus $y_a^\nu < y_b^\nu < y_c^\nu$). 
Models with special flavon directions like the so-called Constrained Sequential Dominance 3 alignment \cite{Bjorkeroth:2016lzs} have simply $\sin \theta_{13} \simeq m_1/(m_1 +m_2)$, which does not constrain the ratio $M_2/M_1$. 
The only constraint on RH neutrino masses comes from Eq.~\eqref{eq:mieqdom}.
Setting $m_1 = m_\mathrm{sol}$ and $m_2 = m_\mathrm{atm}$ implies $M_2\simeq 10^{11}$ GeV and $M_2/M_1 \simeq m_c^2/(6 m_u^2) \simeq 5 \times 10^4$.
Under these conditions $N_1$ contributions to the BAU are far too small, but $N_2$ can still successfully contribute, as shown explicitly in \cite{Bertuzzo:2010et,Blanchet:2011xq,deAnda:2017yeb}.

Unlike this traditional case for $N_2$ leptogenesis, which is typically aimed at resolving the problem of having a lightest neutrino with too small a mass ($M_1 \ll 10^9$ GeV), in our case even the $N_2$ region requires $M_1 \gtrsim 10^9$ GeV, to avoid too-large washout. 
In some sense, separate to the above discussion on naturalness, some balancing is also required to ensure the initial $N_2$ asymmetry, which may be one or two orders of magnitude larger than anticipated by the observed BAU, is washed out just the right amount by $N_1$ interactions to yield the correct value of $Y_B$.

In the UTZ model we can also compare ratios between the $y_{a,b,c}^\nu$ presented here and ratios of their respective counterparts from the up quark sector, which have $y_a^u/y_b^u \sim 0.05$, $y_b^u/y_c^u \sim 0.05^2$ \cite{deMedeirosVarzielas:2017sdv}, in accordance with an expansion parameter $\varepsilon_u \sim 0.05$. 
By contrast, the hierarchy between $y_a^\nu$ and $y_b^\nu$ is only up to one order of magnitude.

In the numerical analysis above, the parameters $\tan \beta$ and $M_c\sim M_3$ are kept fixed. 
The main impact of $\tan\beta$ is only to define the boundary between the two- and three-flavor regimes. 
Given that the model favours large $\tan \beta$ values, we have taken a moderately large value of $\tan \beta = 10$, for which $N_2$ leptogenesis takes place in the two-flavour regime, $N_1$ leptogenesis takes place in the three-flavour regime, and a sizable asymmetry in the electron flavour survives. 
Although the results would be qualitatively similar for larger $\tan \beta$ values, for $\tan \beta < 10$ the entire asymmetry production occurs in the two-flavoured regime and, due to the alignment of $N_1$ and $N_2$ Yukawa couplings in the $e-\mu$ plane, it is more difficult to obtain a sufficient asymmetry. 
Ideally, a more natural realization of $N_2$ leptogenesis would be achieved in the fully three flavoured regime $M_1, M_2 \leq 10^9 (1+\tan\beta^2)$ GeV, but this situation can not be realized while simultaneously maintaining the required hierarchy $M_1 < M_2$. 
Regarding $M_c$, the scenario in which the $N_2$ production dominates is where the ratio $M_2/M_3 $ is large, $M_2/M_3 \gtrsim 0.1$.  
We have considered a value for $M_c$ consistent with the model and which illustrates the relevant leptogenesis features.
Another choice will see the viable $N_2$-leptogenesis region shifted up- or downwards in the mass $M_2$ in order to maintain this large ratio.

\section{Conclusions}
\label{sec:Conclusions}

We have studied the generation of the baryon asymmetry of the Universe through leptogenesis in the Universal Texture Zero $SO(10) \times \Delta(27) \times \mathbb{Z}_N$ flavoured GUT model \cite{deMedeirosVarzielas:2017sdv}.
Here, leptogenesis yields the observed BAU for a considerable region of the parameter space. 
When expressed in terms of the RH neutrino masses $M_1$ and $M_2$, which are functions of the model parameters $M_a$ and $M_b$.
The viable ranges for the mass of the lightest RH neutrino eigenstate have a lower bound of $M_1 \gtrsim 4 \times 10^9$ GeV, which is still barely compatible with a gravitino mass $ m_{3/2} \gtrsim 10$ TeV, provided the gravitino is unstable \cite{Kawasaki:2008qe}.

We specifically considered the effect of $N_2$ leptogenesis, which we conclude to be disfavored: although there exists a region of parameter space where $N_2$ leptogenesis provides the dominant contribution to the final asymmetry, this corresponds to a scenario with both a very strong hierarchy between the two lightest RH neutrinos, i.e. $M_1 \ll M_2$, and comparatively small hierarchy between $M_2$ and $M_3$.
This is not a natural expectation in the model, which predicts a strong hierarchy between the heaviest neutrino and the two lighter ones, i.e. $M_1 < M_2 \ll M_3$.
Lepton asymmetries generated by decays of the heaviest neutrino $N_3$ are therefore also negligible. 

The preferred mechanism is thus $N_1$ leptogenesis.
The requirement that it accounts for the entire baryon asymmetry allows us to restrict the parameters governing the neutrino Yukawa matrix and RH neutrino mass matrix.
These are otherwise only partially constrained by the observed neutrino masses and mixing, namely those combinations of parameters which appear in the neutrino matrix after seesaw.
We find that viable $N_1$ leptogenesis requires $M_1 \gtrsim 4 \times 10^9$ GeV, with $M_2 \gtrsim 2 \times 10^{11}$ GeV, while $0.002 \lesssim M_1/M_2 \lesssim 0.1$. 
By consistency with low energy observables, we can similarly constrain the neutrino Yukawa couplings, which are bounded from below, $y_a^\nu \gtrsim 0.003$, $y_b^\nu \gtrsim 0.008$.

In conclusion, flavoured leptogenesis is viable for the UTZ model in the standard $N_1$ regime.
Through this we are able to place further constraints on the parameter space of the UTZ model, leading to direct constraints on the scale of the parameters $M_a$, $M_b$ governing the RH neutrino masses. 
Given that in the model the active neutrino masses originate from type-I seesaw leading to normal ordering with a strong hierarchy, the leptogenesis constraint on $M_a$, $M_b$ can then be combined with the observed mass-squared differences to indirectly constrain the Dirac neutrino couplings $y^\nu_a$, $y^\nu_b$. 
These constraints are complementary to those provided by the study of flavour-changing processes \cite{deMedeirosVarzielas:2018vab} in the UTZ model.

\section*{Acknowledgements}

IdMV acknowledges funding from the Funda\c{c}\~{a}o para a Ci\^{e}ncia e a Tecnologia (FCT) through the contract IF/00816/2015 and partial support by FCT through projects CFTP-FCT Unit 777 (UID/FIS/00777/2019), CERN/FIS-PAR/0004/2017 and PTDC/FIS-PAR/29436/2017 which are partially funded through POCTI (FEDER), COMPETE, QREN and EU.
AM and OV were supported under MICIU Grant FPA2017-84543-P and by the “Centro de Excelencia Severo Ochoa” programme
under grant SEV-2014-0398. OV acknowledges partial support from the “Generalitat Valenciana”
grant PROMETEO2017-033.
AM acknowledges support from La-Caixa-Severo Ochoa scholarship.

\appendix

\section{Boltzmann equations}
\label{app:boltzmann}

Recall that the baryon asymmetry $ Y_B $ can be expressed as
\begin{equation}
    Y_{B} = \frac{10}{31} \sum_{\fli} Y_{\Delta_\fli} ,
\end{equation}
where $ Y_{\Delta_\fli} $ are the $ B/3 - L_\fli $ asymmetries for each active lepton species $ \fli $. 
In the fully-flavoured scenario, these are simply the usual three lepton flavours, $ \fli = e, \mu, \tau $.
Assuming hierarchical RH neutrinos and thermal leptogenesis, the lepton asymmetries are obtained by solving the Boltzmann equations
\begin{equation}
\begin{aligned}
    \frac{d Y_{N_i}}{dz} &= 
        - 2 D \left(Y_{N_i}-Y_{N_i}^\mathrm{eq}\right) , \\
    \frac{d Y_{\widetilde{N}_{i}}}{dz} &= 
        - 2 D\left(Y_{\widetilde{N}_{i}}-Y_{\widetilde{N}_{i}}^\mathrm{eq}\right) , \\
    \frac{d Y_{\Delta_\fli}}{dz} &= 
         2 \cpasym{\fli}{i} D\left(Y_{N_i}-Y_{N_i}^\mathrm{eq}\right) + 2 \cpasyms{\fli}{i} D \left(Y_{\widetilde{N}_{i}}-Y_{\widetilde{N}_{i}}^\mathrm{eq}\right) 
        + \frac{\decK{\fli}{i}}{\decK{}{i}} W \sum_{\fli^\prime} A_\mathrm{\fli \fli^\prime}Y_{\Delta_{\fli^\prime}} ,
\end{aligned}
\end{equation}
where $z = M_i/T$.
As noted in Section~\ref{sec:Leptogenesis}, the factors $D$ and $W$ govern the decay and washout behaviour.
In this appendix we make these explicit, noting how information about decays and scattering are incorporated.
In particular, we follow \cite{Antusch:2006cw}. 

The equilibrium number density for a given field $f$ is denoted $Y_f^\mathrm{eq}$, and are functions of $z$. 
The RH (s)neutrino densities are given by
\begin{equation}
    Y_{N_i}^\mathrm{eq} = Y_{\widetilde{N}_i}^\mathrm{eq}
        = \frac{45}{2 \pi^4 g_*} z^2 \BesselK{2},
\end{equation}
where $g_* = 228.75$ is the effective number of degrees of freedom in the MSSM and $\BesselK{2}$ the modified Bessel functions of the second kind.
The (s)lepton distributions are given by
\begin{equation}
    Y_{\fli}^\mathrm{eq} = Y_{\widetilde{\fli}}^\mathrm{eq} = \frac{45}{\pi^4 g_*}. 
\end{equation}

We now turn to the decay and washout factors, $D$ and $W$.
There are three classes of processes that contribute to the Boltzmann equations: (1) decays and inverse decays ($N\leftrightarrow L_\fli H_u$), (2) $\Delta L = 1$ scatterings (e.g. $N Q \leftrightarrow L_\fli t$) and (3) $\Delta L=2$ processes ($L_\fli L_\fli \leftrightarrow \overbar{H}_u \overbar{H}_u$, $L_\fli H_u \leftrightarrow \overbar{L}_\fli \overbar{H}_u$). 
Following \cite{Buchmuller:2004nz} and \cite{Antusch:2006cw}, in this analysis we include $\Delta L = 1$ scatterings involving neutrino and top Yukawa couplings, neglecting gauge boson-mediated processes, as well as thermal corrections, and all $\Delta L = 2$ processes.
Then
\begin{equation}
    D = z \decK{}{i} f_1(z) \frac{\BesselK{1}}{\BesselK{2}} , \qquad 
    W =\frac{z}{2} 
    \decK{}{i}
    f_2(z)
    \frac{\BesselK{1}}{\BesselK{2}} 
    \frac{Y_{N_i}^\mathrm{eq}+Y_{\widetilde{N}_i}^\mathrm{eq}}{Y_{\fli}^\mathrm{eq}+Y_{\widetilde{\fli}}^\mathrm{eq}} .
\end{equation}
The effects of $\Delta L = 1$ scatterings are encapsulated in the functions $f_1(z)$ and $f_2(z)$.
These are discussed in \cite{Buchmuller:2004nz} and \cite{Antusch:2006cw}, and we may approximate them by
\begin{equation}
    f_1(z) \simeq 
    f_2(z) \simeq \frac{z}{a} \left[\log \left( 1+\frac{a}{z}\right) + \frac{1}{a \log(M_i/M_h) z} \right] \left( 1+\frac{15}{8 z}\right), 
    \quad
    a = \frac{4 \pi^2 g_{N_i} v_u^2}{9 m_t^2 \log(M_i/M_h)},
\end{equation}
where $M_h=125$ GeV is the Higgs mass, $m_t \simeq 93$ GeV is the top mass (at the GUT scale, which is a reasonable approximation of the value at the leptogenesis scale for our purposes) and $g_{N_i}=2$. 

$A$ is a numerical matrix describing flavour mixing, and is given in the $n_F$-flavour regime by
\begin{equation}
\begin{aligned}
    A = \left\{
        \begin{array}{cc}
            \pmatr{
                -93/110 &  6/55 & 6/55 \\
                3/40 &  -19/30 & 1/30 \\
                3/40 &  1/30 & -19/30
            } , & n_F = 3 \\
            \pmatr{
                -541/761 & 152/761 \\
                46/761 &  -494/761
            } , & n_F = 2 \\
            -1 , & n_F = 1
        \end{array}
    \right. .
\end{aligned}
\label{eq:matrixA}
\end{equation}
While the ratio of Bessel functions, $ \BesselK{1}/\BesselK{2} $, is in principle well-behaved for all positive $z$, for computational efficiency and stability it may be convenient to use the approximations
\begin{equation}
    \BesselK{1} \approx \left(1 + \frac{15}{8 z} \right)^{-1} \BesselK{2} \approx \frac{1}{z} \sqrt{1 + \frac{\pi}{2} z} e^{-z}.
\end{equation}

As described in Section~\ref{sec:Leptogenesis}, the decay factors are given by 
\begin{equation}
    \decK{\fli}{i} = \frac{v_u^2}{m_\ast M_i} (\lambda_\nu^\dagger)_{i \fli}(\lambda_\nu)_{\fli i}, \qquad
    \decK{}{i} = \sum_{\fli} \decK{\fli}{i} ,
\end{equation}
and the CP asymmetries by
\begin{equation}
    \cpasym{\fli}{i} = 
    \cpasyms{\fli}{i} = 
        \frac{1}{8\pi}  \sum_{j \neq i} \frac{
        \mathrm{Im}[(\lambda_\nu^\dagger)_{i\fli} (\lambda_\nu^\dagger \lambda_\nu)_{ij} (\lambda_\nu^T)_{j\fli}]}
        {(\lambda_\nu^\dagger \lambda_\nu)_{ii}}
        g\bigg(\frac{M_j^2}{M_i^2}\bigg) .
\end{equation}
where $g(x)$ is a loop function given by the sum of the vertex and the self energy contributions \cite{DiBari:2005st,Antusch:2006cw}; in the MSSM,
\begin{equation}
    g(x) = \sqrt{x}\left[\frac{2}{1-x}-\log\left(\frac{1+x}{x}\right)\right] .
\end{equation}
In the limit of very hierarchical RH neutrino masses, i.e. $x \ll 1$ or $x \gg 1$, to good approximation,
\begin{equation}
    g(x) = \left\{ \begin{array}{cc}
        -\dfrac{3}{\sqrt{x}}, & x \gg 1 \\
        2 \sqrt{x} \left(1 + \log \sqrt{x} \right) , & x \ll 1 
    \end{array}\right. .
\end{equation}

\section{Seesaw with rank-one matrices}
\label{app:seesawrank1}

We consider here a more intuitive explanation of the texture zero remaining after seesaw, following the method employed for the model in \cite{Bjorkeroth:2015uou}, and presented in \cite{Bjorkeroth:2016lzs}.

The Dirac and Majorana matrices are given in terms of four rank-1 matrices, expressed in terms of the VEV alignments of flavons $ \phi_{a,b,c} $, where
\begin{equation}
    \phi_a \propto (1,1,-1) , \qquad 
    \phi_b \propto (0,1,1) , \qquad 
    \phi_c \propto (0,0,1) .
\end{equation}
For notational simplicity, in tis appendix we use $ \phi_i $ to refer simply to the direction of the VEV (rather than the field or VEV itself).
We have
\begin{equation}
\begin{aligned}
    Y_\nu &= y_a^\nu ( \Box_{ab}+\Box_{ba} )+y_b^\nu \Box_b + y_c^\nu \Box_c , \\
    M_N &= M_a ( \Box_{ab}+\Box_{ba} ) + M_b \Box_b + M_c\Box_c ,
\end{aligned}
\label{eq:Ynu_rank1}
\end{equation}
where
\begin{equation}
    \Box_{ab} = (\Box_{ba})^T = \phi_a \phi_b^T, \qquad
    \Box_b = \phi_b \phi_b^T, \qquad
    \Box_c = \phi_c \phi_c^T.
\end{equation}
We define another set of vectors $ \widetilde{\phi}_{a,b,c} $, which are orthogonal to $ \phi_{a,b,c} $, i.e. 
\begin{equation}
    \widetilde{\phi}_i^T \phi_j = \delta_{ij}, \qquad i,j = a,b,c .
\end{equation}
An appropriate choice is
\begin{equation}
    \widetilde{\phi}_a \propto (1,0,0) , \qquad
    \widetilde{\phi}_b \propto (-1,1,0) , \qquad 
    \widetilde{\phi}_c \propto (2,-1,1) ,
\end{equation}
and new rank-1 matrices $\widetilde{\Box}_{ij} = \widetilde{\phi}_i \widetilde{\phi}_j^T $.
The inverse of the Majorana mass matrix can then be decomposed in terms of the new matrices as
\begin{equation}
    M_N^{-1} = 
        \frac{1}{M_a} (\widetilde{\Box}_{ab} + \widetilde{\Box}_{ba}) 
        - \frac{M_b}{M_a^2} \widetilde{\Box}_a 
        + \frac{1}{M_c} \widetilde{\Box}_c .
\label{eq:invMN_rank1}
\end{equation}

Decomposing $ Y_\nu$ and $ M_N $ as per Eqs.~\eqref{eq:Ynu_rank1} and \eqref{eq:invMN_rank1} and applying the seesaw formula, $m_\nu = -v_u^2 Y_\nu M_N^{-1} Y_\nu^T$, we note that, due the orthogonality of the two sets of vectors, no new matrix structures appear.
In particular, there are no mixed terms with $\phi_c$, e.g. $ \Box_{ac}$, nor a structure $\Box_a$, either of which would spoil the texture zero in $m_\nu$. 
The light neutrino mass matrix in fact preserves the same structure as the other mass matrices in the model, i.e. 
\begin{equation}
    m_\nu = m_a (\Box_{ab} + \Box_{ba}) + m_b \Box_b + m_c \Box_c .
\label{eq:mnu}
\end{equation}
The parameters of the light neutrino mass matrix, $ m_a, m_b, m_c $, are given in terms of $ y^\nu_{a,b,c} $ and $ M_{a,b,c} $ by
\begin{equation}
	m_a = -\frac{v_u^2 {y^\nu_a}^2}{M_a} , \qquad
    m_b = m_a \left(2 \frac{y^\nu_b}{y^\nu_a} - \frac{M_b}{M_a} \right), \qquad
    m_c = -\frac{v_u^2 {y^\nu_c}^2}{M_c} .
\label{app:eq:ma-mc}
\end{equation}
The above discussion holds true also when the first higher-order terms are introduced.
These are given by the superpotential
\begin{equation}
    \delta \mathcal{W}_Y = 
        \overbar{L}_{\nu,i} N_j  H_u \left[\frac{g^\nu_d}{\Lambda^4} (\phi_b^i \phi_c^j + \phi_c^i \phi_b^j) S^2 \right] + \overbar{L}_{\nu,i} N_j H_u \left[\frac{g^\nu_e}{\Lambda^4} (\phi_a^i \phi_c^j + \phi_c^i \phi_a^j) S^2 \right] ,
\end{equation}
and modify Eq.~\eqref{eq:Ynu_rank1} by
\begin{equation}
    \delta Y_\nu = y_d^\nu (\Box_{bc} + \Box_{cb}) + y_e^\nu (\Box_{ac} + \Box_{ca}),
\end{equation}
which preserves the texture zero.
After seesaw, the main effect of these terms can be absorbed into two additional parameters $m_d$ and $m_e$ and in corrections to the relations in Eq.~\eqref{app:eq:ma-mc}. 
Note that the term dependent on $y_e^\nu$ gives rise to a $\Box_a$ structure in $m_\nu$, which spoils the texture zero, but appears only strongly suppressed by a factor $(y_e^\nu)^2/M_c$.

\section{Analytic expression for the leptogenesis parameters}
\label{app:LpGnparam}

In this appendix we give the analytic expression for $\lambda_\nu$, i.e. for the Dirac neutrino matrix in the basis where $Y_\e$ and $M_N$ are diagonal, as well as for the decay factors $\decK{\fli}{i}$ and the $CP$ asymmetries $\cpasym{\fli}{i}$ entering in our numerical calculation. 

The Yukawa and Majorana mass matrices share a unified symmetric and hierarchical texture zero, 
\begin{equation}
    M_{\ell} = \pmatr{
        0 & a & a \\
        a & b+2 a & b\\
        a & b & c+b -2 a
    } , \qquad
    \ell = \e,\nu,N,
\label{eq:Mf}
\end{equation}
where $(a,b,c)$ stand for either $y_{a,b,c}^{\nu,\e}$ or $M_{a,b,c}$. 
Given this complex symmetric structure with the hierarchy $a<b<c$, the diagonalizing matrices satisfy the relation\\
\begin{equation}
    \hat{M}_\ell \equiv V_\ell M_\ell V_\ell^T \approx \diag \left(| a^2/b |, |b|, |c| \right).
\label{eq:MRdiag}
\end{equation}
They can be approximated by
\begin{equation}
    V_\ell = \pmatr{
        1 - \dfrac{1}{2} \left|\dfrac{a}{b}\right|^2 & - \dfrac{a}{b} \left(1 - \dfrac{2 a}{b}\right) & -\dfrac{2a^2}{b c}  \\
        \dfrac{a^*}{b^*} \left(1 - \dfrac{2a^*}{b^*} \right) & 1 - \dfrac{1}{2} \left|\dfrac{a}{b}\right|^2 & - \dfrac{b}{c} \\
       \dfrac{a^*}{c^*} & \dfrac{b^*}{c^*} & 1 
    } 
    P_\ell , 
\label{eq:VP}
\end{equation} 
where $P_\ell = \diag( e^{-i(2 \gamma - \delta + \pi)/2}, e^{-i \delta/2}, 1) $ is a matrix of phases ensuring the eigenvalues in $ \hat{M}_\ell $ are real and positive.
With $ a \sim \epsilon_\nu^3$, $ b \sim \epsilon_\nu^2$, and $ c \sim 1$, Eq.~\eqref{eq:VP} is accurate (and unitary) to $ \mathcal{O}(\epsilon_\nu^3) $, and preserves the (1,1) texture zero to $ \mathcal{O}(\epsilon_\nu^4) $.
Expanding $V_\ell$ as $ V_\ell = (\mathds{1} + \Delta V_\ell)P_\ell$, we split $\lambda_\nu$ into two parts, one which preserves $Y_\nu$ (up to rephasing) and a correction term $\Delta\lambda_\nu^*$, i.e.
\begin{equation}
    \lambda_\nu^* \equiv V_\e Y_\nu V_N^T 
        = P_\e (\mathds{1} + \Delta V_\e) Y_\nu (\mathds{1} + \Delta V_N^T) P_N 
        = P_\e (Y_\nu +\Delta \lambda_\nu^*) P_N 
\label{eq:lambdanu2}
\end{equation}
where
\begin{equation}
    \Delta \lambda_\nu^* \simeq 
    \Delta V_\e Y_\nu + Y_\nu \Delta V_N^T  \simeq -
    \begin{pmatrix}
        y_a^\nu \frac{M_a}{M_b} + y_a^\nu \frac{y_a^\e}{y_b^\e} & y_b^\nu \frac{y_a^\e}{y_b^\e} &  y_b^\nu \frac{y_a^\e}{y_b^\e} \\
        y_b^\nu \frac{M_a}{M_b} & \mathcal{O}\left(y_a^\nu \frac{M_a}{M_b}\right)& y_c^\nu \frac{y_b^\e}{y_c^\e} \\
        y_b^\nu \frac{M_a}{M_b} & \mathcal{O}\left(y_a^\nu \frac{M_a}{M_b}\right) & \mathcal{O}\left(y_b^\nu \frac{y_b^\e}{y_c^\e}\right) 
    \end{pmatrix} ,
\end{equation}
having discarded the term $ \Delta V_\e Y_\nu \Delta V_N^T $.
Notice that the corrections in $\Delta\lambda_\nu^*$ are typically of the order of $Y_\nu$ except in the 22,32 and 33 elements where they are smaller. 
They are never larger; in this sense $\lambda_\nu^*$ preserves the same hierarchy between the elements as $Y_\nu$.
Considering only the first term in Eq.~\eqref{eq:lambdanu2}, $\lambda_\nu^* \sim P_\e Y_\nu P_N$, we can deduce for the decay factors:
\begin{equation}
    \decK{}{1} \sim \frac{v_u^2}{m_*} \left|\frac{{y_a^\nu}^2 M_b}{M_a^2}\right|
    \begin{pmatrix} 0\\1\\1 \end{pmatrix} , \qquad
    \decK{}{2} \sim \frac{v_u^2}{m_*} \left| \frac{{y_b^\nu}^2}{M_b} \right| 
    \begin{pmatrix} \left| \frac{y_a^\nu}{y_b^\nu} \right|^2 \\1\\1 \end{pmatrix} ,
\end{equation}
and for the CP asymmetries,
\begin{equation}
\begin{aligned}
    \cpasym{}{1} &\sim \frac{3}{8\pi} 
    \left| \frac{y_b^\nu M_a}{M_b} \right|^2
    \sin(2 \omega_1) 
    \begin{pmatrix} 0\\1\\1 \end{pmatrix} , \\
    \cpasym{}{2} &\sim 
    \frac{3}{16\pi} \left| \frac{{y_c^\nu} M_b}{M_c} \right|
    \begin{pmatrix} 
        \left| \dfrac{{y_a^\nu}^2}{y_b^\nu} \right| \sin\omega_2 \\
        |y_b^\nu| \sin\omega_2 \\
        |y_c^\nu| \sin(\omega_2 - \delta_\nu)
    \end{pmatrix}
    + \frac{1}{4\pi} \left| 
   \frac{y_a^\nu M_a}{M_b} \right|^2 
    \sin(2 \omega_1) \begin{pmatrix}
        0\\1\\1
    \end{pmatrix} ,
\end{aligned}
\end{equation}
where we have called 
$\omega_1 = [\gamma_N - \gamma_\nu - \delta_N + \delta_\nu]$ and 
$\omega_2 = [\delta_N - \delta_\nu]$ the leading order leptogenesis phases entering the $N_1$ and $N_2$ calculations respectively.
Including also the leading-order contributions from $\Delta\lambda_\nu^*$, we obtain analytic approximations for the decay factors and CP asymmetries in terms of the input parameters of our analysis.
Defining $Q = | M_a y_b^\nu / (M_b y_a^\nu) | $, we have
\begin{equation}
\begin{aligned}
    \decK{}{1} &\simeq \frac{v_u^2}{m_*} \left| \frac{{y_a^\nu}^2 M_b }{M_a^2} \right|
    \left( 1 - 2 Q \cos \omega_1 + Q^2 \right)
    \begin{pmatrix}
        \mathcal{O}\left(\frac{M_a}{M_b}\frac{y_a^\e}{y_b^\e}\right)\\1\\1
    \end{pmatrix} , \\
    \cpasym{}{1} &\simeq \frac{3}{4 \pi} 
    \left| \frac{y_b^\nu M_a}{M_b} \right|^2 
    \frac{\sin \omega_1 (\cos \omega_1 - Q)}{1 - 2 Q \cos \omega_1 + Q^2}
    \begin{pmatrix}
        \mathcal{O}\left(\frac{M_a}{M_b}\frac{y_a^\e}{y_b^\e}\right)\\1\\1
    \end{pmatrix} ,
\end{aligned}
\end{equation}
and, defining $Q_\e=|y_a^\e y_b^\nu / (y_b^\e y_a^\nu)|$ and $\omega_{1,2}^\e=\omega_{1,2}(N\rightarrow\e)$,
\begin{equation}
\begin{aligned}
    \decK{}{2} &\simeq \frac{v_u^2}{m_*} \left| \frac{{y_b^\nu}^2 }{M_b } \right|
    \begin{pmatrix}
        \left| \frac{y_a^\nu}{y_b^\nu} \right|^2 \left(
        1 - 2 Q_\e \cos \omega_1^\e
        + Q_\e^2 
    \right) \\1\\1
    \end{pmatrix} , \\
    \cpasym{}{2} &\simeq \frac{3}{16\pi} \left| \frac{{y_c^\nu} M_b}{M_c} \right|
    \frac{1}{1+2 \,\mathrm{Re}\!\left[\dfrac{y_a^\nu}{y_b^\nu}\right]}\begin{pmatrix} 
        \left| \dfrac{{y_a^\nu}^2}{y_b^\nu} \right| \sin\omega_2(1-2Q_\e \cos\omega_1^\e+Q_\e^2) \\
        |y_b^\nu|\sin\omega_2 - \left|\frac{y_c^\nu y^\e_b}{y_c^\e}\right| \sin(\omega_2+\omega_2^\e)\\
        |y_c^\nu| \sin(\omega_2 - \delta_\nu)
    \end{pmatrix} .
\end{aligned}
\end{equation}


\begin{thebibliography}{99}
\setlength{\parskip}{2pt}
\bibitem{Sakharov:1967dj}
  A.~D.~Sakharov,
  Pisma Zh.\ Eksp.\ Teor.\ Fiz.\  {\bf 5} (1967) 32
   [JETP Lett.\  {\bf 5} (1967) 24]
   [Sov.\ Phys.\ Usp.\  {\bf 34} (1991) no.5,  392]
   [Usp.\ Fiz.\ Nauk {\bf 161} (1991) no.5,  61].

\bibitem{Kuzmin:1985mm}
  V.~A.~Kuzmin, V.~A.~Rubakov and M.~E.~Shaposhnikov,
  Phys.\ Lett.\  {\bf 155B} (1985) 36.

\bibitem{Fukugita:1986hr}
  M.~Fukugita and T.~Yanagida,
  Phys.\ Lett.\ B {\bf 174} (1986) 45.

\bibitem{Blum:2007jz}
  A.~Blum, C.~Hagedorn and M.~Lindner,
  Phys.\ Rev.\ D {\bf 77} (2008) 076004
  [arXiv:0709.3450 [hep-ph]].

\bibitem{Hagedorn:2012pg}
  C.~Hagedorn and D.~Meloni,
  Nucl.\ Phys.\ B {\bf 862} (2012) 691
  [arXiv:1204.0715 [hep-ph]].

\bibitem{Holthausen:2013vba}
  M.~Holthausen and K.~S.~Lim,
  Phys.\ Rev.\ D {\bf 88} (2013) 033018
  [arXiv:1306.4356 [hep-ph]].

\bibitem{Ishimori:2014nxa}
  H.~Ishimori, S.~F.~King, H.~Okada and M.~Tanimoto,
  Phys.\ Lett.\ B {\bf 743} (2015) 172
  [arXiv:1411.5845 [hep-ph]].

\bibitem{Yao:2015dwa}
  C.~Y.~Yao and G.~J.~Ding,
  Phys.\ Rev.\ D {\bf 92} (2015) no.9,  096010
  [arXiv:1505.03798 [hep-ph]].

\bibitem{Varzielas:2016zuo}
  I.~de Medeiros Varzielas, R.~W.~Rasmussen and J.~Talbert,
  Int.\ J.\ Mod.\ Phys.\ A {\bf 32} (2017) no.06n07,  1750047
  [arXiv:1605.03581 [hep-ph]].

\bibitem{Lu:2018oxc}
  J.~N.~Lu and G.~J.~Ding,
  Phys.\ Rev.\ D {\bf 98} (2018) no.5,  055011
  [arXiv:1806.02301 [hep-ph]].

\bibitem{Hagedorn:2018gpw}
  C.~Hagedorn and J.~König,
  arXiv:1811.07750 [hep-ph].

\bibitem{Lu:2019gqp}
  J.~N.~Lu and G.~J.~Ding,
  JHEP {\bf 1903} (2019) 056

\bibitem{deMedeirosVarzielas:2017sdv}
  I.~de Medeiros Varzielas, G.~G.~Ross and J.~Talbert,
  JHEP {\bf 1803} (2018) 007
  [arXiv:1710.01741 [hep-ph]].

\bibitem{deMedeirosVarzielas:2018vab}
  I.~De Medeiros Varzielas, M.~L.~López-Ibáñez, A.~Melis and O.~Vives,
  JHEP {\bf 1809} (2018) 047
  [arXiv:1807.00860 [hep-ph]].

\bibitem{King:2003rf}
  S.~F.~King and G.~G.~Ross,
  Phys.\ Lett.\ B {\bf 574} (2003) 239
  [hep-ph/0307190].

\bibitem{Ross:2004qn}
  G.~G.~Ross, L.~Velasco-Sevilla and O.~Vives,
  Nucl.\ Phys.\ B {\bf 692} (2004) 50
  [hep-ph/0401064].

\bibitem{deMedeirosVarzielas:2005ax}
  I.~de Medeiros Varzielas and G.~G.~Ross,
  Nucl.\ Phys.\ B {\bf 733} (2006) 31
  [hep-ph/0507176].

\bibitem{deMedeirosVarzielas:2011wx}
  I.~de Medeiros Varzielas,
  JHEP {\bf 1201} (2012) 097
  [arXiv:1111.3952 [hep-ph]].

\bibitem{Varzielas:2012ss}
  I.~de Medeiros Varzielas and G.~G.~Ross,
  JHEP {\bf 1212} (2012) 041
  [arXiv:1203.6636 [hep-ph]].

\bibitem{Bjorkeroth:2015uou}
  F.~Björkeroth, F.~J.~de Anda, I.~de Medeiros Varzielas and S.~F.~King,
  Phys.\ Rev.\ D {\bf 94} (2016) no.1,  016006
  [arXiv:1512.00850 [hep-ph]].

\bibitem{Bjorkeroth:2015tsa}
  F.~Björkeroth, F.~J.~de Anda, I.~de Medeiros Varzielas and S.~F.~King,
  JHEP {\bf 1510} (2015) 104
  [arXiv:1505.05504 [hep-ph]].

\bibitem{Varzielas:2015aua}
  I.~de Medeiros Varzielas,
  JHEP {\bf 1508} (2015) 157
  [arXiv:1507.00338 [hep-ph]].

\bibitem{Gatto:1968ss}
  R.~Gatto, G.~Sartori and M.~Tonin,
  Phys.\ Lett.\  {\bf 28B} (1968) 128.

\bibitem{Antusch:2006cw}
  S.~Antusch, S.~F.~King and A.~Riotto,
  JCAP {\bf 0611} (2006) 011
  [hep-ph/0609038].

\bibitem{Davidson:2002qv}
  S.~Davidson and A.~Ibarra,
  Phys.\ Lett.\ B {\bf 535} (2002) 25
  [hep-ph/0202239].

\bibitem{Giudice:2003jh}
  G.~F.~Giudice, A.~Notari, M.~Raidal, A.~Riotto and A.~Strumia,
  Nucl.\ Phys.\ B {\bf 685} (2004) 89
  [hep-ph/0310123].

\bibitem{Buchmuller:2004nz}
  W.~Buchmuller, P.~Di Bari and M.~Plumacher,
  Annals Phys.\  {\bf 315} (2005) 305
  [hep-ph/0401240].

\bibitem{Abada:2006fw}
  A.~Abada, S.~Davidson, F.~X.~Josse-Michaux, M.~Losada and A.~Riotto,
  JCAP {\bf 0604} (2006) 004
  [hep-ph/0601083].

\bibitem{DiBari:2005st}
  P.~Di Bari,
  Nucl.\ Phys.\ B {\bf 727} (2005) 318
  [hep-ph/0502082].

\bibitem{DiBari:2010ux}
  P.~Di Bari and A.~Riotto,
  JCAP {\bf 1104} (2011) 037
  [arXiv:1012.2343 [hep-ph]].

\bibitem{DiBari:2013qja}
  P.~Di Bari and L.~Marzola,
  Nucl.\ Phys.\ B {\bf 877} (2013) 719
  [arXiv:1308.1107 [hep-ph]].

\bibitem{DiBari:2014eya}
  P.~Di Bari, L.~Marzola and M.~Re Fiorentin,
  Nucl.\ Phys.\ B {\bf 893} (2015) 122
  [arXiv:1411.5478 [hep-ph]].

\bibitem{DiBari:2014eqa}
  P.~Di Bari, S.~King and M.~Re Fiorentin,
  JCAP {\bf 1403} (2014) 050
  [arXiv:1401.6185 [hep-ph]].

\bibitem{Bjorkeroth:2016lzs}
  F.~Björkeroth, F.~J.~de Anda, I.~de Medeiros Varzielas and S.~F.~King,
  JHEP {\bf 1701} (2017) 077
  [arXiv:1609.05837 [hep-ph]].

\bibitem{Kawasaki:2008qe}
  M.~Kawasaki, K.~Kohri, T.~Moroi and A.~Yotsuyanagi,
  Phys.\ Rev.\ D {\bf 78} (2008) 065011
  [arXiv:0804.3745 [hep-ph]].

\bibitem{Jedamzik:2006xz}
  K.~Jedamzik,
  Phys.\ Rev.\ D {\bf 74} (2006) 103509
  [hep-ph/0604251].

\bibitem{Pradler:2006qh}
  J.~Pradler and F.~D.~Steffen,
  Phys.\ Rev.\ D {\bf 75} (2007) 023509
  [hep-ph/0608344].

\bibitem{Das:2016czs}
  D.~Das, M.~L.~López-Ibáñez, M.~J.~Pérez and O.~Vives,
  Phys.\ Rev.\ D {\bf 95} (2017) no.3,  035001
  [arXiv:1607.06827 [hep-ph]].

\bibitem{Lopez-Ibanez:2017xxw}
  M.~L.~López-Ibáñez, A.~Melis, M.~J.~Pérez and O.~Vives,
  JHEP {\bf 1711} (2017) 162
   Erratum: [JHEP {\bf 1804} (2018) 015]
  [arXiv:1710.02593 [hep-ph]].

\bibitem{Lopez-Ibanez:2019rgb}
  M.~L.~López-Ibáñez, A.~Melis, D.~Meloni and Ó.~Vives,

\bibitem{Barbieri:1999ma}
  R.~Barbieri, P.~Creminelli, A.~Strumia and N.~Tetradis,
  Nucl.\ Phys.\ B {\bf 575} (2000) 61
  [hep-ph/9911315].

\bibitem{Nardi:2006fx}
  E.~Nardi, Y.~Nir, E.~Roulet and J.~Racker,
  JHEP {\bf 0601} (2006) 164
  [hep-ph/0601084].

\bibitem{Antusch:2010ms}
  S.~Antusch, P.~Di Bari, D.~A.~Jones and S.~F.~King,
  Nucl.\ Phys.\ B {\bf 856} (2012) 180
  [arXiv:1003.5132 [hep-ph]].

\bibitem{Vives:2005ra}
  O.~Vives,
  Phys.\ Rev.\ D {\bf 73} (2006) 073006
  [hep-ph/0512160].

\bibitem{DiBari:2015oca}
  P.~Di Bari and S.~F.~King,
  JCAP {\bf 1510} (2015) no.10,  008
  [arXiv:1507.06431 [hep-ph]].

\bibitem{Bertuzzo:2010et}
  E.~Bertuzzo, P.~Di Bari and L.~Marzola,
  Nucl.\ Phys.\ B {\bf 849} (2011) 521
  [arXiv:1007.1641 [hep-ph]].

\bibitem{Blanchet:2011xq}
  S.~Blanchet, P.~Di Bari, D.~A.~Jones and L.~Marzola,
  JCAP {\bf 1301} (2013) 041
  [arXiv:1112.4528 [hep-ph]].

\bibitem{deAnda:2017yeb}
  F.~J.~de Anda, S.~F.~King and E.~Perdomo,
  JHEP {\bf 1712} (2017) 075
   Erratum: [JHEP {\bf 1904} (2019) 069]
  [arXiv:1710.03229 [hep-ph]].
\end{thebibliography}
\end{document}